\documentclass[a4paper,fleqn]{cas-dc}
\usepackage{enumitem}
\usepackage[authoryear]{natbib}
\usepackage[figuresright]{rotating}
\usepackage{moreverb}
\usepackage{amsmath,amssymb,amsfonts}
\usepackage{algorithmic}
\usepackage{url}
\usepackage{graphicx}
\usepackage{textcomp}
\usepackage{xcolor}
\usepackage{tcolorbox}
\usepackage{ragged2e}
\usepackage{balance}
\usepackage{colortbl}
\usepackage{multirow}
\usepackage{color}

\usepackage{CJKutf8}
\usepackage{siunitx}
\usepackage{rotfloat}
\usepackage{float}
\usepackage{lscape}
\usepackage{amssymb,mathrsfs,amsmath}
\usepackage{adjustbox}
\def\tsc#1{\csdef{#1}{\textsc{\lowercase{#1}}\xspace}}
\tsc{WGM}
\tsc{QE}
\tsc{EP}
\tsc{PMS}
\tsc{BEC}
\tsc{DE}
\begin{document}
\begin{sloppypar}

\let\WriteBookmarks\relax
\def\floatpagepagefraction{1}
\def\textpagefraction{.001}
\shorttitle{Issues, Causes and Solutions of LLM Open-Source Projects}
\shortauthors{Y Cai et al.}
\title [mode = title]{Demystifying Issues, Causes and Solutions in LLM Open-Source Projects}

\author[1]{Yangxiao Cai}
\ead{yangxiaocai@whu.edu.cn}
\credit{Conceptualization, Investigation, Data curation, Formal analysis, Writing - Original draft preparation}

\author[1]{Peng Liang}
\cormark[1]
\ead{liangp@whu.edu.cn}
\credit{Conceptualization, Methodology, Investigation, Data curation, Supervision, Writing - Original draft preparation}
\address[1]{School of Computer Science, Wuhan University, China}

\author[1]{Yifei Wang}
\ead{whiten@whu.edu.cn}
\credit{Investigation, Data curation, Writing - Original draft preparation}

\author[2]{Zengyang Li}
\ead{zengyangli@ccnu.edu.cn}
\credit{Conceptualization, Methodology, Writing - review and editing}
\address[2]{School of Computer Science \& Hubei Provincial Key Laboratory of Artificial Intelligence and Smart Learning, Central China Normal University, China}

\author[3]{Mojtaba Shahin}
\ead{mojtaba.shahin@rmit.edu.au}
\credit{Conceptualization, Methodology, Writing - review and editing}
\address[3]{School of Computing Technologies, RMIT University, Australia}

\cortext[cor1]{Corresponding author.}

%\fundinginfo{Natural Science Foundation of Hubei
%Province of China, Grant Number: 2021CFB577\\
%National Natural Science Foundation of China, Grant Number: 62176099}

\begin{abstract}
With the advancements of Large Language Models (LLMs), an increasing number of open-source software projects are using LLMs as their core functional component. Although research and practice on LLMs are capturing considerable interest, no dedicated studies explored the challenges faced by practitioners of LLM open-source projects, the causes of these challenges, and potential solutions. To fill this research gap, we conducted an empirical study to understand the issues that practitioners encounter when developing and using LLM open-source software, the possible causes of these issues, and potential solutions. We collected all closed issues from 15 LLM open-source projects and labelled issues that met our requirements. We then randomly selected 994 issues from the labelled issues as the sample for data extraction and analysis to understand the prevalent issues, their underlying causes, and potential solutions. Our study results show that (1) \textit{Model Issue} is the most common issue faced by practitioners, (2) \textit{Model Problem}, \textit{Configuration and Connection Problem}, and \textit{Feature and Method Problem} are identified as the most frequent causes of the issues, and (3) \textit{Optimize Model} is the predominant solution to the issues. Based on the study results, we provide implications for practitioners and researchers of LLM open-source projects.
\end{abstract}

\begin{keywords}
Large Language Model, GitHub, Open Source Project, Issue, Cause, Solution  
\end{keywords}

\maketitle

\section{Introduction}
\label{sec:introduction}

%textit{The Emergence of LLMs:}
With the advancements of Pre-trained Language Models (PLMs) in recent years, there has been a significant breakthrough in the capacity of Language Models (LMs) to handle very large-scale data, which has led to the emergence of Large Language Models (LLMs) \citep{hou2024large,zhao2023survey}. LLMs refer to neural network language models trained on massive text datasets with billions of parameters. The most advanced LLMs to date have demonstrated remarkable language comprehension and generation capabilities \citep{zheng2023empirical}.

%{The Research of LLMs:}
%With outstanding natural language processing capabilities, LLMs have received significant academic interest among practitioners.
Since the release of ChatGPT in 2022 \citep{chatgpt}, there has been a noticeable increase in research related to LLM \citep{zhao2023survey}. The Software Engineering (SE) research community has also extensively used LLMs as tools to solve SE tasks. This is evident in recent review papers on LLM in SE. For example, Zheng \textit{et al.} collected relevant literature on LLMs from seven literature databases, categorizing these papers according to the SE tasks involved, then reviewed the current research status of LLMs from the perspective of the seven major tasks in software development \citep{zheng2023empirical}. In addition, Hou \textit{et al.} conducted a systematic literature review on the application of LLMs in SE \citep{hou2024large}. They investigated which LLMs were utilized for SE tasks, and the collection and preprocessing of SE-related datasets for these models. They also investigated strategies for optimizing and evaluating LLM performance in SE, and the SE tasks LLMs successfully addressed.

%{The Practice of LLMs in open-source projects:}
Conversely, software practitioners are increasingly integrating LLMs as components within their software systems \citep{weber2024large, liu2023prompt}. These systems, known as ``LLM-based Systems'', utilize the capabilities of LLMs to perform tasks that typically require substantial coding effort \citep{weber2024large}. This trend is also evident in open-source software (OSS) projects, primarily due to the emergence of open-source LLMs and frameworks. For example, LangChain, an open-source software framework for building applications based on LLMs, is currently receiving significant attention from practitioners \citep{langchain}. In addition, Microsoft's semantic-kernel, an SDK that integrates LLMs into popular programming languages such as C\#, Python, and Java, has also received widespread attention \citep{semantic-kernel}. In this paper, we refer to OSS projects that leverage LLMs as \textbf{LLM open-source projects}.

%{Github Issue Collecting:}
Despite the increasing number of LLM open-source projects, there is no systematic research on the issues experienced in LLM open-source project development, the causes of the issues, and potential solutions from the perspective of practitioners. This study aims to address this gap by studying approximately 1000 GitHub closed issue discussions from LLM open-source projects. Considering that GitHub is presently the largest hosting platform for open-source projects in the world, and numerous related studies have utilized community data from open-source software for empirical software engineering work, we have decided to select LLM open-source projects from GitHub.
%To ensure the acquisition of pertinent information, such as the causes of issues and the solutions to these issues, to the fullest extent possible, we have determined to select the closed issues of every repository for our analysis. Finally, adhering to this principle, we have chosen appropriate projects from GitHub and extracted closed issues from them as research samples to complete our study.

\textbf{Our findings show that}: (1) \textit{Model Issue} is the most common issue faced by practitioners, (2) \textit{Model Problem}, \textit{Configuration and Connection Problem}, and \textit{Feature and Method Problem} are identified as the most frequent causes of the issues, and (3) \textit{Optimize Model} is the predominant solution to the issues.

The \textbf{main contributions} of this work: (1) We identified the issues occurred in the development and utilization of LLM open-source software with the dataset~\citep{dataset} and provided a two-level taxonomy for these issues. (2) We identified the causes and solutions of these issues, and came up with a taxonomy for these causes and solutions. 
(3) We provided the mapping relationship between the identified issues and causes, as well as the mapping relationship between the issues and solutions. %and (4) We discussed the implications based on the study results, highlighting areas that practitioners and researchers should pay more attention to.

The rest of this paper is structured as follows: Section \ref{sec:methodology} presents the Research Questions (RQs) and the research process employed in this study. Section \ref{sec:results} presents the study results with their interpretation. Section \ref{sec:implication} discusses the implications based on the study results. Section \ref{sec:threats} outlines the potential threats to validity. Section \ref{sec:relatedWork} reviews the related work, and Section \ref{sec:conclusions} concludes this work along with the directions for future research.

\section{Methodology}
\label{sec:methodology}
\begin{figure*}[htbp]
	\centering
	\includegraphics[width=1\textwidth]{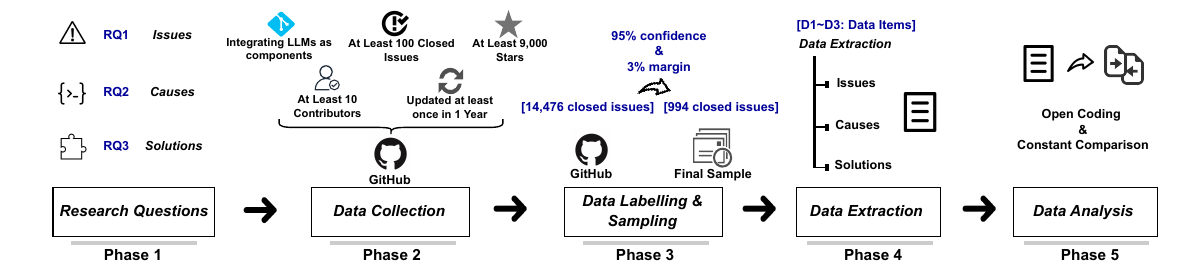}
	\caption{\textcolor{black}{Overview of the research process}}
	\label{fig:Overview of research process}
\end{figure*}

Our study aims to identify the issues in developing and using LLM open-source software and to determine the causes of these issues as well as possible solutions. With this aim in mind, we formulate three RQs to guide the subsequent phases of the methodology, as shown in Fig.~\ref{fig:Overview of research process}. The RQs and their rationales are detailed in Section~\ref{Research Questions}.

\subsection{Research Questions} 
\label{Research Questions}
% Five RQs are formulated according to the goal of this study:\\
\noindent \textbf{RQ1: What issues do OSS practitioners encounter when developing and using LLM open-source software?}\\
\noindent \emph{Rationale}: As mentioned in Section \ref{sec:introduction}, remarkable advances in LLMs \citep{zhao2023survey} have led to a growing number of open-source projects leveraging LLMs as one of their core functional components (i.e., LLM open-source projects). The answer of this RQ can provide an overall picture of the issues that OSS developers experience when developing LLM open-source projects. It also opens new directions for subsequent research on how these issues can be resolved.
%\noindent \emph{Rationale}: As discussed in Section \ref{sec:introduction}, LLM open-source projects have been widely used in research and practice. Since the release of ChatGPT in 2022, there has been a remarkable advance in research related to LLMs, with the quantity of studies steadily increasing \citep{zhao2023survey}. Therefore, the answer of this RQ can help practitioners discover the issues faced by LLM open-source projects currently. It also points out the direction for our subsequent research on the causes of issues and solutions to address these issues.

\noindent \textbf{RQ2: What are the underlying causes of these issues?}\\
\noindent \emph{Rationale}: After collecting the issues currently faced by practitioners as identified in RQ1, it is important to further identify the causes of these issues. By identifying the causes, the study can provide insights for finding solutions and inspire subsequent refinement and optimization of both LLM open-source software and the LLMs.
%The answer of this RQ will play a significant role in our search for solutions (RQ3) and in the subsequent refinement and optimization of both LLM open-source software and the LLMs.

\noindent \textbf{RQ3: What are the current and potential solutions to address these issues?}\\
\noindent \emph{Rationale}: This RQ aims to understand how practitioners address the issues they encounter. The answer to this RQ can identify the most commonly used solutions, and provide various potential solutions from practitioners to the issues collected in RQ1. Ultimately, by exploring existing solutions to these issues, the study can provide guidance for the optimization of the software, thereby enhancing its performance.

\subsection{Data Collection}
We collected the data for our study from GitHub, which is currently the largest platform for hosting open-source software projects. In addition, every project on GitHub has an ``Issues'' page, which provides a platform for developers and users of the project to raise issues encountered while developing and using the software, and to discuss the identification, tracing, and resolution of these issues with other developers and users. \textcolor{black}{In order to thoroughly explore the causes of issues and their potential solutions, we decided to collect all ``closed'' issues from the selected LLM open-source project repositories.}

To collect data as comprehensively as possible, we did not set a start date for the source of the data. The date we conducted the data search is December 11, 2023, and we collected the data that were created before this date. The first author utilized a keyword search approach, connecting the search terms ``LLM'' and ``Large Language Model'' with an ``OR'' logic, and then searched for projects on the entire GitHub. Any project that contains the search terms (e.g., project description, tags) was retrieved.

After obtaining the search results, we found that although many projects include ``LLM'' or ``Large Language Model'' terms, they do not use LLMs as components to implement their functionality. For example, some project are to teach LLMs without using LLMs in the projects. 
%We expected that the projects should be open-source software projects explicitly based on LLMs and sufficiently representative, which indicates that these projects should have enough popularity. Therefore, we used the number of stars to sort the search results in descending order. 
Therefore, we set the following criteria for selecting LLM open-source projects:

\begin{enumerate}[label=(\arabic*)] 
    \item The project must be a real open-source software, which indicates that the project must be a real open-source software project that integrates LLMs as components (e.g., APIs) to implement its main functionality. They should not be a course teaching project, teaching guide project, LLM tutorial, LLM-related books, examination or competition related to LLMs, or other projects that merely use our keyword in project documentation without employing an LLM framework or functionality. 
    \item The project must have at least 100 closed issues.
    \item The project must have at least 9,000 stars.
    \item The project must have at least 10 contributors. 
    \item The project must have been updated at least once within the past year.
\end{enumerate}

In our criteria, we defined criterion (1) to filter projects that utilize LLMs as components to implement their main functionality \cite{malavolta2021mining}. Similar to other empirical studies (e.g., \cite{waseem2023empirical,zheng2024realistic}) that employed project information (e.g., star count) as criteria for selecting appropriate GitHub projects with sufficient popularity and representativeness, we defined criteria (2), (3), (4), and (5) to exclude unpopular, inactive, or non-representative LLM open-source projects.

\begin{table}[htbp]
\scriptsize
\caption{Information of selected LLM open-source projects}
\label{Selected Projects}
\begin{tabular}{>{\raggedright}m{2.4cm}m{0.71cm}<{\centering}m{1.32cm}<{\centering}m{0.71cm}<{\centering}m{0.71cm}<{\centering}}
\hline
\textbf{Project Name}   & \textbf{\#Issues}      & \textbf{\#Contributors}  & \textbf{\#Forks}   & \textbf{\#Stars}      \\ \hline
langchain               & 3475                   & 1991                     & 10.6k              & 71.9k                 \\ \hline
gpt4all                 & 901                    & 79                       & 6.1k               & 57k                   \\ \hline
MetaGPT                 & 97                     & 45                       & 3.7k               & 32.5k                 \\ \hline
FastChat                & 919                    & 206                      & 3.7k               & 30.3k                 \\ \hline
text-generation-webui   & 2561                   & 264                      & 3.9k               & 29.8k                 \\ \hline
ollama                  & 552                    & 89                       & 1.1k               & 27.6k                 \\ \hline
quivr                   & 875                    & 95                       & 3k                 & 26.4k                 \\ \hline
autogen                 & 223                    & 359                      & 2.2k               & 18.8k                 \\ \hline
ChatDev                 & 149                    & 40                       & 2.1k               & 18.5k                 \\ \hline
unilm                   & 745                    & 54                       & 2.2k               & 16.7k                 \\ \hline
semantic-kernel         & 1030                   & 195                      & 2.2k               & 15.6k                 \\ \hline
SuperAGI                & 269                    & 61                       & 1.6k               & 13.4k                 \\ \hline
haystack                & 2381                   & 232                      & 1.5k               & 12k                   \\ \hline
dolly                   & 157                    & 13                       & 1.2k               & 10.0k                 \\ \hline
pandas-ai               & 254                    & 59                       & 782                & 9.3k                  \\ \hline
\end{tabular}
\end{table}

Following the selection criteria, the first author ultimately selected 15 LLM open-source projects (see Table~\ref{Selected Projects}) that met our criteria. \textcolor{black}{The domains of the 15 selected LLM open-source projects are categorized into five categories (see Table~\ref{Domains of Selected Projects}).} Then, the first author employed web scrapers to collect all the closed issues from these project repositories. In the end, we gathered a total of 14,588 closed issues. Each closed issue was recorded in an MS Excel file. We named each MS Excel file using the convention: `\textsc{Project Name}' + `\textsc{\_}' + `\textsc{Issue Number}'.

\begin{table}[htbp]
\scriptsize
\caption{\textcolor{black}{Domain categorization of the selected LLM open-source projects}}
\label{Domains of Selected Projects}
\begin{tabular}{>{\raggedright}m{1.42cm}m{2.9cm}m{2.43cm}<{\centering}}
\hline
\textbf{Domain}                        & \textbf{\#Definition}                                                                    & \textbf{\#Projects}               
                                    \\ \hline
LLM Development Framework               & Open-source projects that provide frameworks for developing software by leveraging LLMs   & langchain, MetaGPT, quivr, autogen, ChatDev, semantic-kernel, SuperAGI, haystack, dolly  \\ \hline
LLM Running Tool                        & Open-source projects for running LLMs in specific environments                           & gpt4all, FastChat, ollama             
                                    \\ \hline
LLM Data Analysis Tool                  & Open-source projects for data analysis using LLMs                                         & pandas-ai                             
                                    \\ \hline
LLM Training Tool                       & Open-source projects that support pre-training across tasks, languages, and modalities    & unilm                                 
                                    \\ \hline
LLM Frontend Tool                       & Open-source projects that develop user interface for LLMs                                 & text-generation-webui                 
                                    \\ \hline
\end{tabular}
\end{table}

\subsection{Data Labelling and Sampling}
After collecting closed issues from the 15 LLM open-source projects, \textcolor{black}{we found that some issues raised by practitioners may not be related to the problems with the projects. For example, some practitioners only raised an issue to chat with others or to learn the features of GitHub, like a practitioner simply submitted an issue titled ``\textit{hi}'' as a greeting, without any content related to the project. In addition, not all issues have corresponding category labels (e.g., pandas-ai issue \#85). Relying solely on existing labels could lead to the omission of critical information. Therefore, }we conducted data labelling to identify issues suitable for our study. The closed issues we expected should contain specific information related to the issues that practitioners encountered while developing or using LLM open-source software. \textcolor{black}{For example, a user mentioned in text-generation-webui \#2926 that ``\textit{the model is loaded into memory without any errors, but crashes on generation of text}'', which explicitly indicates an issue present in the LLM open-source project text-generation-webui.} Then, on the basis of the \textcolor{black}{pilot and formal} labelling results (as detailed in Section~\ref{pilotdatalabelling} and Section~\ref{formaldatalabelling}, respectively), we selected a representative data sample (as detailed in Section~\ref{datasampling}) to construct the dataset for data extraction. 

\subsubsection{Pilot Data Labelling}\label{pilotdatalabelling}
To minimize potential personal biases, the first and third authors independently conducted a pilot data labelling. We randomly selected 50 closed issues from all closed issues, and then the first and third authors independently labelled them. The two authors labelled the data that contain information related to issues arising during the development or use of LLM open-source software. The inter-rater consistency on data labelling results between the two authors was measured using the Cohen's Kappa coefficient~\citep{jacob1960coefficient}, yielding a value of 0.838, which indicates a \textcolor{black}{very high} level of data labelling consistency between the two authors. For any disagreements or conflicts in the pilot data labelling results, the two authors discussed together with the second author to reach a consensus. Finally, wo got 32 issues out of 50 issues are considered relevant. The results of pilot data labelling were compiled and recorded in MS Excel files~\citep{dataset}.

\subsubsection{Formal Data Labelling}\label{formaldatalabelling}
The first author then conducted the formal data labelling. In this process, we excluded the issues not related to our study. The projects we selected utilized LLMs to achieve the main functions of the software. We only kept those issues related to the problems that practitioners encountered during the development and use of LLM open-source software. Ultimately, the first author collected a total of 14,476 closed issues. The results of formal data labelling were compiled and recorded in MS Excel files~\citep{dataset}.

\subsubsection{Data Sampling}\label{datasampling}
\begin{figure}[htbp]
	\centering
	\includegraphics[width=\linewidth]{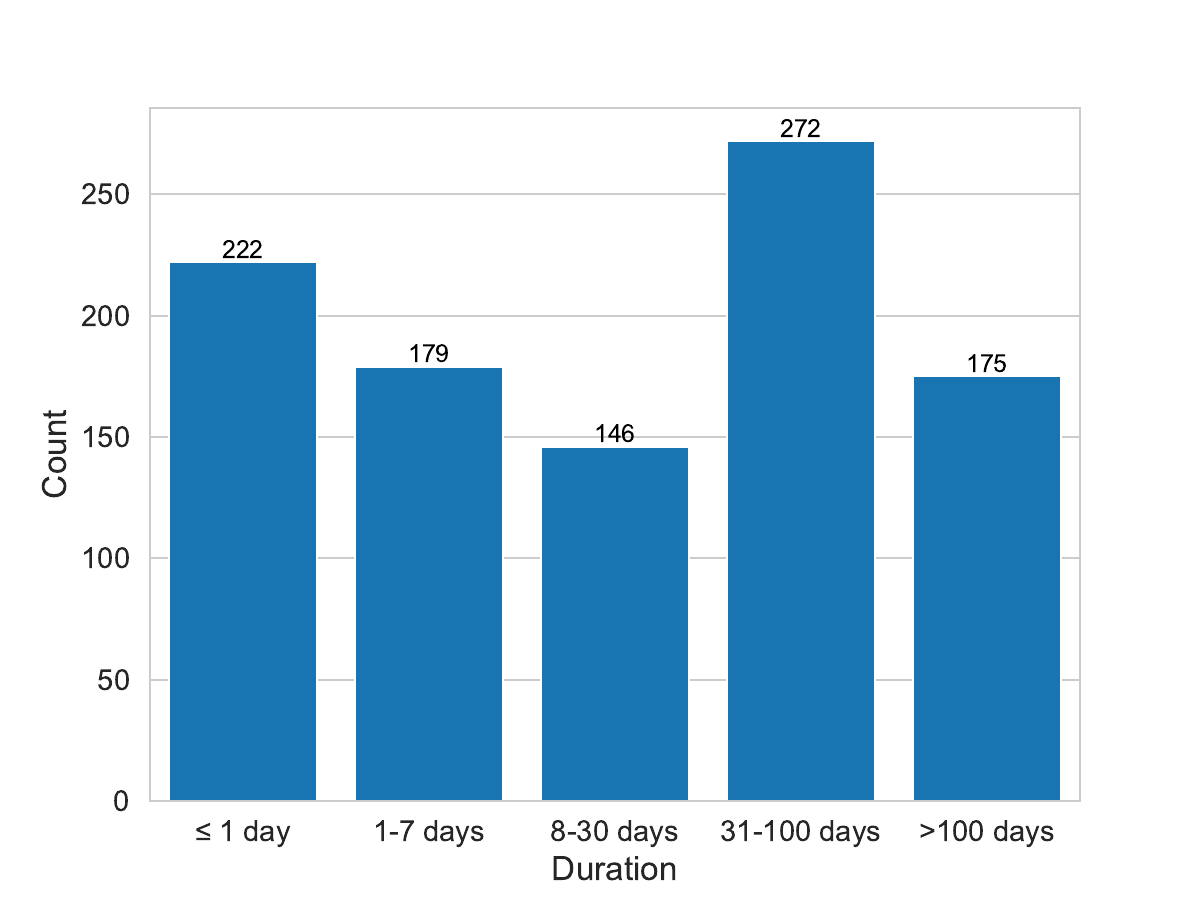}
	\caption{\textcolor{black}{The distribution of the duration of the closed issues in the final data sample}}
	\label{fig:The distribution of the duration of closed issues in the final sample}
\end{figure}

We found that the dataset comprising a total of 14,476 closed issues was too large for manual data extraction. Therefore, we decided to select a suitable number of closed issues. We randomly selected 994 closed issues to form a representative set as our dataset for data extraction with a 95\% confidence level and a 3\% margin of error~\citep{israel1992determining}. \textcolor{black}{The source projects of the closed issues in the final data sample are presented in Table~\ref{Sources of Issues in Final Sample}. In addition, the distribution of the duration from issue opening to closing for the closed issues in the final data sample is shown in Fig.~\ref{fig:The distribution of the duration of closed issues in the final sample}.}

\begin{table}[htbp]
\scriptsize
\caption{\textcolor{black}{The source projects of the closed issues in the final data sample}}
\label{Sources of Issues in Final Sample}
\begin{tabular}{>{\raggedright}m{2.4cm}m{1.05cm}<{\centering}m{1.32cm}<{\centering}m{1.42cm}<{\centering}}
\hline
\textbf{Project Name}   & \textbf{\#Selected Issues}      & \textbf{\#Contributors}  & \textbf{\#Lines of Code}      \\ \hline
langchain               & 226                             & 1991                     & 1,468.7k                      \\ \hline
gpt4all                 & 50                              & 79                       & 238.1k                        \\ \hline
MetaGPT                 & 9                               & 45                       & 133.1k                        \\ \hline
FastChat                & 63                              & 206                      & 121.7k                        \\ \hline
text-generation-webui   & 178                             & 264                      & 162.9k                        \\ \hline
ollama                  & 49                              & 89                       & 276.4k                        \\ \hline
quivr                   & 52                              & 95                       & 61.4k                         \\ \hline
autogen                 & 19                              & 359                      & 1,037.4k                      \\ \hline
ChatDev                 & 11                              & 40                       & 1,372k                        \\ \hline
unilm                   & 50                              & 54                       & 2,047.9k                      \\ \hline
semantic-kernel         & 75                              & 195                      & 726.4k                        \\ \hline
SuperAGI                & 22                              & 61                       & 62.1k                         \\ \hline
haystack                & 162                             & 232                      & 140.3k                        \\ \hline
dolly                   & 9                               & 13                       & 1.6k                          \\ \hline
pandas-ai               & 19                              & 59                       & 84.0k                         \\ \hline
\textbf{Total}          & \textbf{994}                                                                               \\ \hline
\end{tabular}
\end{table}

%we determined the dataset capacity for the formal data extraction should be calculated based on the standard of a 95\% confidence level and a 3\% margin of error. Ultimately, it was concluded that we needed to select 994 issues from a total of 14,476 closed issues to form our dataset. Then, 
%In order to ensure the reliability and authenticity of the research and analysis results as much as possible, and to ensure that the selected data for the dataset is sufficiently representative, we determined the dataset capacity should be calculated according to the standard of a 95\% confidence level and a 3\% margin of error~\citep{israel1992determining}. Ultimately, it was concluded that we need to select 994 issues from a total of 14,476 closed issues to form our dataset. Then, we randomly selected 994 closed issues, and the first author independently conducted the formal data labelling. 

\subsection{Data Extraction} 
To answer the three RQs presented in Section \ref{Research Questions}, we defined a set of data items for data extraction, as shown in Table~\ref{Data Items and RQ}. The data items D1-D3 are respectively used to extract information about issues, underlying causes, and solutions that address these issues, corresponding to RQ1-RQ3. These three data items can be extracted from any part of the content of the collected closed issues, including the title, description, and discussion of the issue. \textcolor{black}{To ensure the accuracy of the data extraction results and the validity of subsequent data analysis, we decided to conduct the data extraction manually. The other reason is that the data sample size was not very large, and manual data extraction was feasible. Therefore, the data extraction was conducted manually.}

\subsubsection{Pilot Data Extraction}
We randomly selected 50 closed issues from our dataset. Next, the first author independently conducted the pilot data extraction on the 50 issues. The results of pilot data extraction indicated that all three data items (see Table~\ref{Data Items and RQ}) could be extracted from our dataset.
Based on the pilot data extraction results, we reached a consensus and defined the following rules for formal data extraction: 
(1) In principle, only one \textit{issue} can be extracted from each closed issue. If there are multiple issues that can be extracted, we record the first issue discussed.
(2) If there are multiple \textit{causes} mentioned in a closed issue, we record all the causes \textcolor{black}{of the issue we extract in the former step, which are }identified by the reporter and the development team.
%There may be more than one \textit{cause} for an \textit{issue}, thus multiple causes can be identified, any cause identified for contributing to the issue will be chosen by us.
(3) If there are multiple \textit{solutions} mentioned in a closed issue, we record all the solutions identified by the reporter and the development team \textcolor{black}{that can address the issue we extract in the former step}.
%There may be more than one \textit{solution} for an \textit{issue}, thus multiple solutions can be identified, any solution that is confirmed to have resolved the issue will be chosen by us.
%We only extracted \textit{solutions} that explicitly addressed an \textit{issue}. If there were multiple solutions to the issue, the first and the second authors should reach a consensus before extracting.

\subsubsection{Formal Data Extraction}
\label{sec:formalDataExtraction}
After completing the pilot data extraction, the first author independently conducted the formal data extraction on the dataset to extract the data based on the data items defined in Table~\ref{Data Items and RQ}. If there were any questions during the data extraction process, the first author discussed these questions with the second author to resolve the doubts. After the first author completed the formal data extraction, the second author reviewed the extraction results and then discussed the results with the first author to reach a consensus on the inconsistencies. The first and second authors conducted multiple rounds of reviews and revisions on the data extraction results to obtain the final results. The data extraction results were compiled and recorded in an MS Excel file~\citep{dataset}.

\textcolor{black}{In the formal data extraction phase, we totally got 11 closed issues that discussed multiple issues in one closed issue in the final data sample. However, according to our criteria, we only kept the first issue discussed and its corresponding causes and solutions.}

\textcolor{black}{Besides, n}ot all closed issues had a corresponding \textit{cause} and \textit{solution} which could be extracted. For some closed issues, the discussion content of the issues was too vague to discern specific information (i.e., \textit{cause} or \textit{solution}) that could be extracted. Moreover, if an issue of a project is not discussed or updated for a long period, it will be automatically closed and turn into a closed issue, making it impossible to extract valid \textit{cause} and \textit{solution}. An issue may also be temporarily closed due to its low priority for resolution. Similarly, an issue might be too complex to identify its causes and solutions. %In addition, failure to identify or determine the cause of an issue can lead to the failure of extracting the cause.%the cause of an issue had not been identified, making it difficult to extract its cause. %or its cause indeterminable, also leading to extraction failure. 
%In summary, each closed issue was quite different from the others. Therefore, we only extracted valuable information to answer the three RQs with the constraints of the defined criteria.

\subsection{Data Analysis}
After completing the data extraction, we conducted data analysis to answer the three RQs formulated in Section \ref{Research Questions}. We employed Open Coding and Constant Comparison methods for data analysis, which are commonly used methodologies within Grounded Theory for qualitative data analysis in software engineering research \cite{stol2016grounded}. 
In Grounded Theory, Open Coding refers to the initial process of data analysis. In this process, researchers meticulously check the text to identify crucial concepts and categories, assigning descriptive codes to these elements. Constant Comparison is an approach used to refine and integrate these concepts and categories. In this process, researchers employ a bottom-up methodology to recurrently compare new data with previously coded data to refine and adjust existing categories based on their differences and similarities.

The steps in our data analysis are as follows: 
(1) The first author meticulously read through the text of each closed issue, then assigned descriptive codes that succinctly summarize the core themes according to the three RQs. For example, the issue of langchain \#2174 was coded as ``\textit{Model Test Failure}'', which indicates that an error occurred while testing the model leading to failure. 
(2) The first author compared the commonalities and differences between codes. Following the principles of a bottom-up approach, the first author categorized identical or similar codes, and used consistent descriptive language to describe the higher-level types and categories. For example, the code from langchain \#2174, together with other similar codes, constitutes the \textsc{Model Test Issue} type, which further belongs to \textit{Model Issue} category.
(3) Once uncertainties about the code descriptions arose, the first author engaged in discussions with the second author to reach a consensus. Due to the nature of Constant Comparison, the initial results underwent several iterations and modifications before reaching the final results.
(4) The second author reviewed and verified the initial analysis results. In case of any doubts or disagreements, the first and the second authors engaged in discussions to reach a final consensus and ultimately resolved the conflict.
\textcolor{black}{After resolving 31 conflicts in the categories of issues, 22 conflicts in the categories of causes, and 27 conflicts in the categories of solutions, we obtained the final results of the data analysis.}

\begin{table}[htbp]
\scriptsize
\caption{Data items extracted and their corresponding RQs}
\label{Data Items and RQ}
\begin{tabular}{m{0.1cm}<{\centering}m{1.2cm}m{5.1cm}m{0.3cm}<{\centering}}
\hline
\textbf{\#}  & \textbf{Data Item}    & \textbf{Description}                                                     
& \textbf{RQ}   \\ \hline
D1          & Issue                 & \textit{The most critical issue discussed in a GitHub issue. In principle, only one issue can be extracted from every closed issue.}           
&  RQ1   \\ \hline
D2          & Cause                 & \textit{The cause(s) for the issue is clearly clarified. Due to the complexity of some issues, there may have multiple causes.}             
&  RQ2   \\ \hline
D3          & Solution              & \textit{The solutions(s) that can solve the issue is clearly provided. Some issues may have multiple solutions.}          
&  RQ3   \\ \hline
\end{tabular}
\end{table}

\section{Results and Interpretation}
\label{sec:results}
In this section, we report the study results of the three RQs and provide their interpretation. 

For the \textit{Issue}, \textit{Cause} and \textit{Solution} taxonomies, we provide a two-tier classification, i.e., categories and types. We also provide brief descriptions of the types under each category. To show the relationship between issues and their causes and solutions, we present the mapping relationships between the \textit{Issue} categories and the \textit{Cause} categories, as well as between the \textit{Issue} categories and the \textit{Solution} categories. It should be noted that only closed issues from which a cause has been extracted are represented in the issue-cause mapping. Similarly, only closed issues from which a solution has been extracted are represented in the issue-solution mapping. 

\begin{figure*}[htbp]
	\centering
	\includegraphics[width=\linewidth]{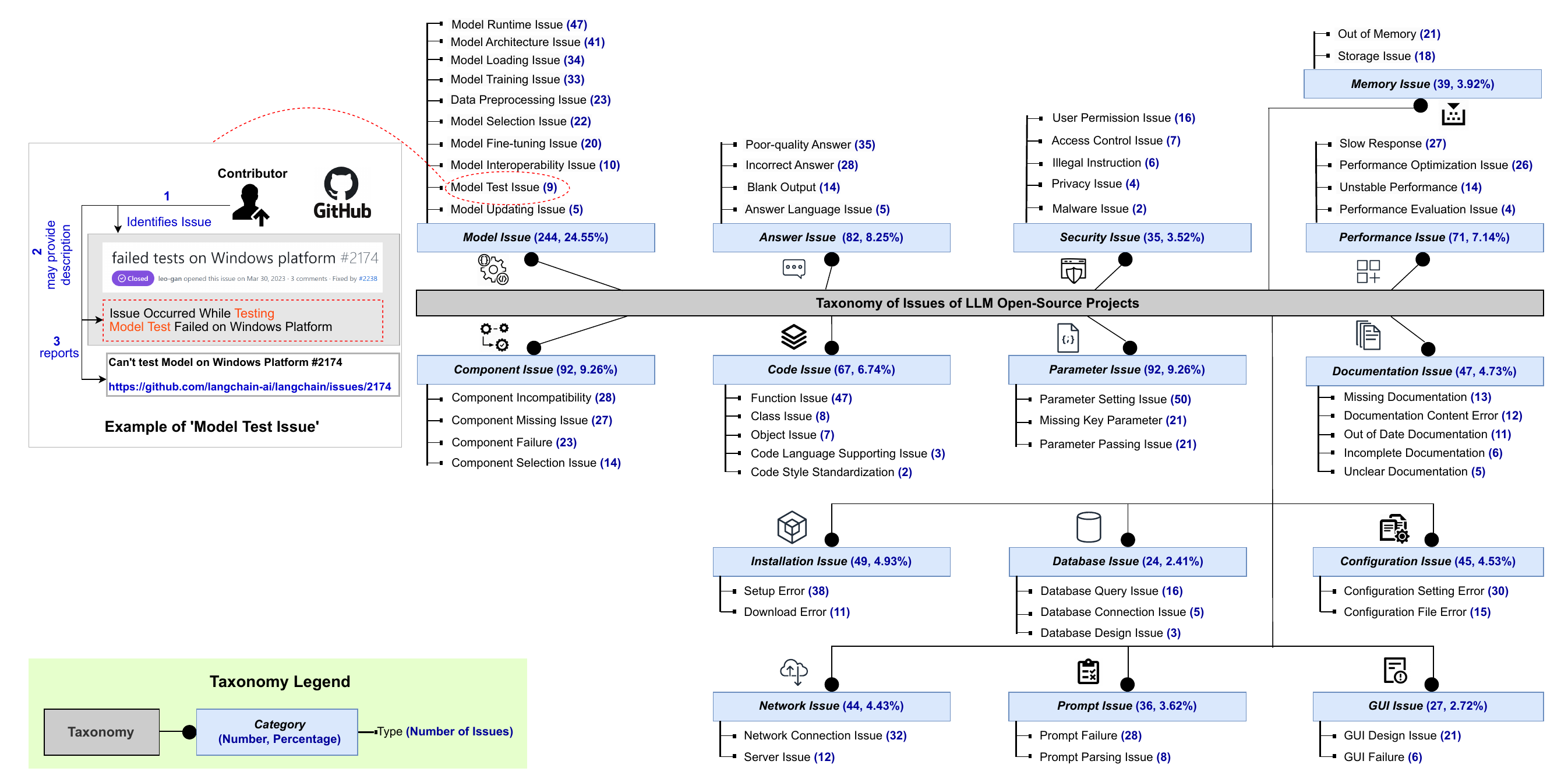}
	\caption{Taxonomy of issues of LLM open-source projects}
	\label{fig:Results of RQ1}
\end{figure*}

\subsection{Category of Issues (RQ1)}
\label{subsec:issuesresults}

Fig.~\ref{fig:Results of RQ1} presents the taxonomy of the issues extracted from the dataset. It can be observed that \textit{Model Issue} (24.55\%) accounts for the majority of issues encountered by practitioners of LLM open-source projects. In addition, a substantial number of practitioners have raised \textit{Component Issue} (9.26\%) and \textit{Parameter Issue} (9.26\%). There is also a portion of practitioners who have encountered \textit{Answer Issue} (8.25\%), which indicates that they did not get answers they want from LLM open-source software, while smaller percentages (less than 8\%) were identified for \textit{Performance Issue} (7.14\%), \textit{Code Issue} (6.74\%),  \textit{Installation Issue} (4.93\%), \textit{Documentation Issue} (4.73\%),  \textit{Configuration Issue} (4.53\%), \textit{Network Issue} (4.43\%), \textit{Memory Issue} (3.92\%), \textit{Prompt Issue} (3.62\%), \textit{Security Issue} (3.52\%), \textit{GUI Issue} (2.72\%), and \textit{Database Issue} (2.41\%).

\subsubsection{\textit{Model Issue (24.55\%)}} \textit{Model Issue} refers to issues related to the models that practitioners encounter while developing and using LLM open-source software. This category encompasses the following ten types.
\begin{itemize}
    \item  \textsc{Model Runtime Issue} refers to the issues that arise during the runtime of LLMs. The core functions of LLM open-source software are performed by the integrated models. Therefore, anomalies or crashes in the models during runtime have a significant impact on the operation of LLM open-source software. For example, a user found that ``\textit{the model is loaded into memory without any errors, but crashes on generation of text}'' (text-generation-webui \#2926).
    
    \item \textsc{Model Architecture Issue} refers to issues related to the design and implementation of LLM architecture. Issues in this type are primarily manifested in unreasonable architectural design (e.g., layers) and unsuitable implementation of mechanisms (e.g., model training), leading to anomalies or failures in the training and use of LLMs. For example, a user received a message that ``\textit{InstructorEmbedding is not found}'', which is a component used to form the embedding layer of the model (langchain \#867).
    
    \item  \textsc{Model Loading Issue} refers to the issues that arise during the loading of LLMs, such as loading failures, incomplete loading, loading the wrong model. Once the model fails to load, the subsequent functions of  LLM open-source software cannot proceed normally. For example, a user reported that ``\textit{an error was thrown when loading with Exllama or Exllamav2 even though pip indicates they are installed}'' (text-generation-webui \#4293).
    
    \item  \textsc{Model Training Issue} refers to the issues that arise during the training of LLMs. These issues include inefficient training algorithms, the need for optimization of model training algorithms, sudden interruptions or crashes during training. For example, a user asked ``\textit{is DEiT-3 trained from scratch only based on the BEiT initialization algorithm or using the similar stagewise pre-training strategy introduced in VLMo}'', which indicates that doubts regarding the model training algorithm exist (unilm \#1022).
    
    \item  \textsc{Data Preprocessing Issue} refers to the issues that arise during data preprocessing in LLMs, including issues in data segmentation, normalization, embedding, and other related processes. The quality of the data received by the models significantly affects the results of model training and operating. For example, a user proposed to the development team that the text splitters could be enhanced to ``\textit{try and generate text fragments}'', which can aid users in better locating and citing specific portions of text (langchain \#5451).
    
    \item  \textsc{Model Selection Issue} refers to issues concerning choosing proper LLMs to implement required functionalities. Many LLM open-source projects support a variety of LLMs. Therefore, selecting one or multiple appropriate models to implement functions and accomplish tasks is crucial. For example, a user asked ``\textit{which trained model to choose for GPU-12GB, Ryzen 5500, 64GB}'', which indicates an issue of selecting models (gpt4all \#324).
    
    \item \textsc{Model Fine-tuning Issue} refers to the issues occurring when fine-tuning LLMs in order to adapt the models for different functions or software integration. To implement new features or incorporate a new model into the software, it is necessary to fine-tune the current models to complete the aforementioned tasks. However, fine-tuning LLMs may potentially lead to issues. For example, a user asked ``\textit{is there any example how to use it (or other scripts) to fine-tune this model}'' (FastChat \#1159).
    
    \item \textsc{Model Collaboration Issue} refers to the issues arising when multiple LLMs collaborate to complete tasks. When accomplishing certain tasks (e.g., information retrieval), it is necessary for several models to work together and interact with each other. However, issues such as incompatibility between models may occur. For example, a user reported that ``\textit{default models should be published when Haystack 1.19.0 is released, to prevent incompatibility}'' (haystack \#5321).
    
    \item \textsc{Model Test Issue} refers to the issues that arise during the testing of LLMs. When testing the models, issues may occur such as test failures, inconsistent or inaccurate test results, and incomplete or missing testing procedures. For example, a user ``\textit{got a timeout error}'' when testing an LLM called GPT-JT (langchain \#648).
    
    \item \textsc{Model Updating Issue} refers to the issues occurring during the model update process, which include update failures, incorrect update versions, update interruptions, and anomalies caused by updates. For example, a user reported that ``\textit{the new OpenAI client library released this week breaks the MT-bench evaluation script}'', which indicates that updating to the new version of the OpenAI client library caused the existing script to malfunction (FastChat \#2657).
\end{itemize}

\textbf{Interpretation}: As a core functional component of LLM open-source software, practitioners are more concerned with functionality and performance of the model, making \textit{Model Issue} be the most prevalent category of issues in developing and using LLM open-source software. \textsc{Model Runtime Issue} (47) and \textsc{Model Architecture Issue} (41) are the top two types of \textit{Model Issue}. \textsc{Model Runtime Issue} highlights the instability of LLMs at runtime and high demands for the deployment environment of LLM open-source software, while \textit{Model Architecture Issue} indicates unreasonable connections between layers and unsuitable implementation of training mechanisms.
%Addressing these two most prominent issues, we believe that significant improvements are still needed in the architecture design and operational stability of LLMs. 
 
\subsubsection{\textit{Component Issue (9.26\%)}} \textit{Component Issue} refers to the issues that arise within or between the components of LLM open-source software. These issues may affect user experience and stability of LLM open-source software. This category encompasses the following four types.
\begin{itemize}
    \item  \textsc{Component Incompatibility} refers to incompatibility issues between components of LLM open-source software, such as version incompatibility, incompatibility in data types, or even functional incompatibility. For example, a user reported that ``\textit{function calls do not work with Azure OpenAI currently due to differences between the OpenAI API and AzureOpenAI}'' (autogen \#78).
    
    \item  \textsc{Component Missing Issue} refers to the need for additional or extended components of LLM open-source software. The emergence of new requirements that necessitate a new component, or the realization that an essential function lacks a component for implementation. For example, to ``\textit{implement real-world, persistent Document Store for v2}'', a user suggested that the haystack team integrate a component called ``\textit{ElasticSearch 8}'', which is an open-source search engine (haystack \#5326).
    
    \item  \textsc{Component Failure} refers to the malfunctiones and crashes of components. Such failures constitute a significant issue affecting the functionality and the user experience of LLM open-source software. For example, a user reported that the API ``\textit{crashed on the second response}'' when using a model named ggml-vicuna-7b (text-generation-webui \#816).
    
    \item  \textsc{Component Selection Issue} refers to the issues of choosing the right components in the development of LLM open-source software. When implementing the specific features, the selection of components is crucial. For example, a user requested the langchain team to support a ``\textit{multi-core loader}'', because the loader called \textit{DirectoryLoader} was ``\textit{using a single core and missing the opportunity of leverage the multiple cores}'' (langchain \#3720).
\end{itemize}

\textbf{Interpretation}: \textit{Component Issue} arises from the failure to use appropriate components to implement the corresponding functionality, or even the absence of required components. We found that under the \textit{Component Issue} category, the top two types: \textsc{Component Incompatibility} (28) and \textsc{Component Missing Issue} (25), have similar proportions. It shows the poor compatibility between components developed by different developers and development communities. Additionally, older components may become incompatible with newer versions.
%It shows that components originating from different developers or communities can lead to . Additionally, older components may become incompatible with newer versions, leading to these two types of issues.
%These three types of issues have a significant impact on \textit{Component Issues}, which indicates that the compatibility, scalability, and stability of LLM open-source components still require optimization. 

\subsubsection{\textit{Parameter Issue (9.26\%)}} \textit{Parameter Issue} refers to issues concerning the parameters of LLM open-source software. The definition, input, passing, and output of parameters are critical steps in the process of the development and use of LLM open-source software. The fact that \textit{Parameter Issue} ranks third among all categories of issues emphasizes its importance. This category encompasses the following three types.
\begin{itemize}
    \item \textsc{Parameter Setting Issue} refers to the issues with the settings of parameters within LLM open-source software. Improper parameter settings may result in poor performance or even anomalies. The settings of parameters include not only the values of the parameters but also their formats, names, length constraints, and the definitions of their data structures. Ensuring reasonableness and consistency of parameter settings is crucial for facilitating subsequent passing and calculation. For example, a user suggested the langchain team that parameters, such as ``\textit{OPENAI\_BASE\_URL, OPENAI\_API\_KEY, CLAUDE2\_BASE\_URL, and CLAUDE2\_API\_KEY}'', can be replaced to increase flexibility in handling the integration and use of other APIs (ChatDev \#6).
    
    \item \textsc{Missing Key Parameter} refers to the issues arising from the absence of certain parameters. There is complex logic for parameter passing and calculation within LLMs. The lack of key parameters can significantly impact the training and effectiveness of the model. For example, a user reported that ``\textit{the parameter `sources' is empty but it should be included in a list called `result' as a string}'' (langchain \#5536).
    
    \item \textsc{Parameter Passing Issue} refers to the issues that occur during the process of parameter passing in LLM open-source software. These issues manifest in aspects such as type conversion of parameters, the composition of parameter sets, and data formatting during the passing process. For example, a user was ``\textit{unable to provide `llm\_chain' to initialize\_agent() while initializing the agent}'' (langchain \#4437).
\end{itemize}

\textbf{Interpretation}: \textit{Parameter Issue} is a critical concern in the development and use of LLM open-source software. LLMs need to handle a large number of parameters, which presents significant challenges for developers. \textsc{Parameter Setting Issue} (50) is the largest type of \textit{Parameter Issue}, which indicates that there are issues with unreasonable parameter values, inappropriate parameter types, missing parameters, and redundant parameters in the development and use of LLM open-source software. Moreover, the complexity of large-scale parameters also presents a significant challenge for developers.
%\textsc{Parameter Setting Issue} (50), which is the largest type of \textit{Parameter Issue}, arises from the improper initial configuration of parameters.

\subsubsection{\textit{Answer Issue (8.25\%)}} \textit{Answer Issue} refers to issues concerning the responses provided by LLM open-source software to the questions given by users. The effectiveness of the responses of LLM open-source software is a crucial characteristic that reflects the user experience of the software. This category encompasses the following four types.
\begin{itemize}
    \item \textsc{Poor-quality Answer} refers to responses produced by LLM open-source software that are correct but low quality, such as being overly verbose with excessive repetition, failing to generate responses according to the specified formats provided by users, and providing responses that are too brief and lack details. For example, when a user asked ``\textit{a few questions in a conversation, there is no context between questions}'', which indicates that the memory of the models is too short to give satisfactory answers (ollama \#8).
    
    \item \textsc{Incorrect Answer} refers to responses where wrong answers are provided by LLM open-source software, such as responses that are not pertinent to the question asked, factually inaccurate, or illogical. For example, a user reported that ``\textit{the results both from Azure OpenAI and from OpenAI are really random and have nothing to do with prompts}'', which indicates that answers are given randomly (semantic-kernel \#1337).
    
    \item \textsc{Blank Output} refers to issues that LLM open-source software fails to provide a response, resulting in an empty answer, or is unable to successfully deliver a response. For example, a use reported that the answers given by the model ``\textit{are all blank}'' (FastChat \#2319).
    
    \item  \textsc{Answer Language Issue} refers to the inadequacy in the variety of human natural languages supported by the LLM open-source software. If the software could support a broader range of human languages, it would significantly enhance user experience. For example, a user asked the gpt4all team to ``\textit{support Chinese model}'' (gpt4all \#1199).
\end{itemize}

\textbf{Interpretation}: \textit{Answer Issue} is a kind of quality issue related to answers generated by LLM open-source software. It is evident that the most prominent concerns in this category are \textsc{Poor-quality Answers} (35) and \textsc{Incorrect Answers} (28). \textit{Answer Issue} arises due to the limitations of LLMs, which are unable to provide satisfactory answers. \textit{Answer Issue} is also related to the way users formulate their questions. Some users do not follow the template specified by the development team when posing their questions, leading to responses that do not meet their expectations.
%There are methods (e.g., organizing prompts using the training data format) available to improve the quality of the answers generated by LLM open-source software.

\subsubsection{\textit{Performance Issue (7.14\%)}} \textit{Performance Issue} refers to issues concerning the performance of LLM open-source software. This category encompasses the following four types.
\begin{itemize}
    \item \textsc{Slow Response} refers to the issues in which LLM open-source software shows a delayed response to requests. The response time significantly impacts the user experience. Sometimes, practitioners even encounter timeouts when waiting for answers. For example, a user reported that the application, which enables users to run and manage LLMs on their local machines, ``\textit{got stuck}'' when running it and had to ``\textit{quit manually}'' (ollama \#1382).
    
    \item \textsc{Performance Optimization Issue} refers to the discussions on how to enhance the performance of LLM open-source software. Performance optimization of LLM open-source software has been a concern of practitioners. For example, one user reported that ``\textit{our deployment gives only 50 English words in 6 seconds}'', and the user inquired with the development team about how to optimize the output token per second (FastChat \#1041).
    
    \item \textsc{Unstable Performance} refers to the instability of the performance of LLM open-source software under varying conditions. The performance of LLM open-source software may deteriorate due to longer execution time, or significant performance discrepancies may arise when execute on different devices or software environments. For example, a user reported that the software ``\textit{started lagging when it got past 3 lines and can take up to a minute to complete}'', which indicates that the response time of the software decreased over time (text-generation-webui \#1542).
    
    \item \textsc{Performance Evaluation Issue} refers to the discussions on how to evaluate the performance of LLM open-source software. Practitioners attempt to discuss and establish a unified set of evaluation criteria to measure the performance of LLM open-source software. For example, a user wanted to know the performance of the model ``\textit{compared to vicuna-13b-v1.3 model}'' and the ranking of this model in the Hugging Face (HF) community, which reflects concerns of users about the performance evaluation of the model (FastChat \#2152).
\end{itemize}

\textbf{Interpretation}: \textit{Performance Issue} mainly stems from the inadequacies and instability of the model performance. Due to \textsc{Slow Response} (27) and \textsc{Unstable Performance} (14) of LLM open-source software, \textsc{Performance Optimization Issue} (26) and \textsc{Performance Evaluation Issue} (4) have been great concerns of practitioners. This category of issues reflect dissatisfaction of users with the response speed and performance stability of LLM open-source software.

\subsubsection{\textit{Code Issue (6.74\%)}} \textit{Code Issue} refers to the programming issues of LLM open-source software. This category encompasses the following five types.
\begin{itemize}
    \item  \textsc{Function Issue} refers to issues concerning the implementation of functions of LLM open-source software, such as the definition of the parameter list of the function, the naming of the function, issues within the code logic of the function, code syntax errors. For example, a user reported that the method ``\textit{`max\_marginal\_relevance\_search()' was not implemented}'', which led to the failure of word embedding (langchain \#10059).
    
    \item  \textsc{Class Issue} refers to issues concerning the implementation of classes of LLM open-source software, which covers issues about methods and parameters of a class, as well as defining new classes for specific requirements. For example, a user complained that ``\textit{it is impossible for anything outside of Microsoft.SemanticKernel.Abstractions to actually implement this abstraction}'', and asked the development team to modify the class ``\textit{IKernel}'', in order to improve the data extraction abilities of models (semantic-kernel \#3195).
    
    \item  \textsc{Object Issue} refers to the issues with the instantiation and management of objects within LLM open-source software, which includes issues related to the manipulation of existing objects, management of object life cycles. For example, a user reported that ``\textit{property `id\_hash\_keys' of the `Document' objects cannot be set}'', which led to the failure of metadata comparison (haystack \#1920).
    
    \item \textsc{Code Language Supporting Issue} refers to requirements to support more programming languages in LLM open-source software. For example, a user asked the langchain team ``\textit{to support the kotlin language}'' in order to ensure ``\textit{the safety and reliability of the software}'', which is an open-source framework for building LLM applications (langchain \#4963).
    
    \item \textsc{Code Style Standardization} refers to the discussion of normalizing coding styles. For example, a user suggested to ``\textit{standardize our code}'' of the software, which is used for building end-to-end question-answering systems and retrieval-augmented generation models (haystack \#2022).
\end{itemize}

\textbf{Interpretation}: \textit{Code Issue} highlights the programming problems within LLM open-source projects. These concerns range from the definition and implementation of functions to issues in object-oriented programming and problems with code style standardization. \textsc{Function Issue} (47) is the most common type of \textit{Code Issue}, as LLM open-source software involves complex algorithms, making the implementation and maintenance of functions challenging for developers.
%Among these types of issues, \textsc{Function Issue} (47) is the most prominent.

\subsubsection{\textit{Installation Issue (4.93\%)}} \textit{Installation Issue} refers to the issues that arise during the installation of LLM open-source software. This category encompasses the following two types.
\begin{itemize}
    \item \textsc{Setup Error} refers to the issues encountered by practitioners during the installation or initialization process of LLM open-source software. For example, a user reported that the software, which is a localized open-source LLM, was ``\textit{not working on Windows after fresh installation}'' (gpt4all \#400).
    
    \item \textsc{Download Error} refers to the issues that arises during the process of downloading LLM open-source software. For example, a user reported that ``\textit{the download speed falls to 100 KB or something}'' after 5\% when downloading the localized LLM (gpt4all \#734).
\end{itemize}
\textbf{Interpretation}: \textit{Installation Issue} reflects the errors and anomalies that occur during the download and initialization of LLM open-source software. LLM open-source projects require compatibility with specific hardware, operating systems, and machine learning libraries. \textsc{Setup Error} (38) as the dominant type of \textit{Installation Issue} highlights the demands that LLM open-source software places on the deployment environment.
%If the system environment does not meet these requirements, \textit{Installation Issue} may occur, which also contributes to \textsc{Setup Error} (38) being the dominant type of \textit{Installation Issue}.

\subsubsection{\textit{Documentation Issue (4.73\%)}} \textit{Documentation Issue} refers to issues with the writing and management of documentation related to LLM open-source projects. This category encompasses the following five types.
\begin{itemize}
    \item \textsc{Missing Documentation} reflects a lack of relevant documentation in LLM open-source projects, which leads to barriers for practitioners in developing and using the software. For example, a user suggested the FastChat team ``\textit{adding tutorials to increase token limits for self-deployment}'' in order to let the model be able to generate longer answers (FastChat \#638).
    
    \item \textsc{Documentation Content Error} reflects that some parts of the documentation of LLM open-source projects contain inaccuracies or errors. Documentation content does not align with reality or there are inconsistencies among different parts of the documentation. For example, a user posted the screenshots of two documents that introduce the evaluation results of the models and asked ``\textit{which one is correct}'', showing inconsistency between the two documents (FastChat \#1782).
    
    \item \textsc{Out of Date Documentation} reflects that certain portions of the documentation of LLM open-source projects are outdated. It shows that during the processes of software development, some documents have not been updated, resulting in outdated documentation which may lead to confusion among practitioners. For example, a user reported that ``\textit{documentation linked in official article is outdated}'', which contains information on model testing (semantic-kernel \#4038).
    
    \item \textsc{Incomplete Documentation} reflects that certain parts of the documentation of LLM open-source projects are not complete. For example, a user asked, ``\textit{can you also suggest how/what to modify this model to run on a single/double GPU as I have access to those}'', which indicates that key information is missing from the documentation of the LLM open-source project (dolly \#8).
    
    \item \textsc{Unclear Documentation} refers to the situation that certain documentation of LLM open-source projects provides ambiguous or obscure descriptions. Unclear descriptions lead to comprehension challenges, creating difficulties for practitioners. For example, a user was quite confused about the documentation related to the model parameters and asked ``\textit{could you please make this more descriptive}'' (autogen \#728).
\end{itemize}
\textbf{Interpretation}: \textit{Documentation Issue} covers the issues in the documentation of LLM open-source projects, which lead to challenges in understanding, developing, and using of LLM open-source software for practitioners. \textsc{Missing Documentation} (13), \textsc{Documentation Content Error} (12), and \textsc{Out of Date Documentation} (11) are the top three types of \textit{Documentation Issue}. This category of issues reflects that developers do not regularly review and update the documentation for LLM open-source projects.

\subsubsection{\textit{Configuration Issue (4.53\%)}} \textit{Configuration Issue} refers to issues concerning the configuration of LLM open-source software. This category encompasses the following two types.
\begin{itemize}
    \item \textsc{Configuration Setting Error} refers to the issues arising from setup settings of LLM open-source projects, including environmental variables, system configurations, component configurations, and machine learning library configurations. For example, a user received a message that ``\textit{Ray Dependency Conflict}'' while configuring a distributed computing framework named Ray (haystack \#1657).
    
    \item \textsc{Configuration File Error} refers to issues concerning configuration files of LLM open-source projects, including missing configuration files, reading and loading errors of configuration files. For example, a user was ``\textit{unable to gather the deployment name from the .env file}'', which led to the failure of deploying the localized model (semantic-kernel \#1474).
\end{itemize}
\textbf{Interpretation}: \textit{Configuration Issue} covers issues related to configuration encountered by practitioners during the deployment of LLM open-source software. LLM open-source software involves complex configuration options, leading to \textsc{Configuration Setting Error} (30) during the configuration process. Moreover, if the project configuration tutorials are not sufficiently detailed or the guidance is unclear, users are more likely to make mistakes or fail to let LLMs meet their requirements when configuring LLM settings.
%\textit{Configuration Issue} indicates that the configuration of software is also a contributing factor to problems in LLM open-source projects. \textsc{Configuration Setting Error} (30) is the largest type of \textit{Configuration Issue}. \textsc{Configuration Setting Error} arises because users do not carefully read or fully understand the project configuration tutorials.

\subsubsection{\textit{Network Issue (4.43\%)}} \textit{Network Issue} refers to network environment issues concerning LLM open-source software, including issues with network connections and servers. This category encompasses the following two types.
\begin{itemize}
    \item \textsc{Network Connection Issue} refers to the errors concerning network connections when using LLM open-source software, including network connection interruptions, connection timeouts, connection errors, and configuration issues related to network connection. For example, a user ``\textit{cannot connect to remote host}'' of an open-source search engine, when performing word embeddings (langchain \#7995).
    
    \item \textsc{Server Issue} refers to the issues in the servers of LLM open-source software, including server connection anomalies, server out of service, and server optimization problems. For example, a user reported a ``\textit{langchain-server error}'' when preprocessing dataset of the model (langchain \#822).
\end{itemize}
\textbf{Interpretation}: \textit{Network Issue} covers the network-related issues and failures encountered by practitioners during the installation and use of LLM open-source software. \textit{Network Issue} can affect the operation of models and responses to requests. \textsc{Network Connection Issue} (32) reflects how unstable network environments and network restrictions prevent LLM open-source software from properly responding to user requests. \textsc{Server Issue} (12) highlights the insufficient performance of servers hosting LLM open-source software, their inadequate load-handling capacity when faced with heavy requests, and poor maintenance of server configurations (e.g., firewall rules).

\subsubsection{\textit{Memory Issue (3.92\%)}} \textit{Memory Issue} refers to issues concerning memory management and the data storage in memory when using LLM open-source software. This category encompasses the following two types.
\begin{itemize}
    \item \textsc{Out of Memory} refers to the situation where available memory is exhausted and no additional memory can be allocated for LLM open-source software, which indicates that the computational and storage resources required to run LLM open-source software are substantial. For example, a user received an error message that ``\textit{OutOfMemoryError: CUDA out of memory. -- train dolly v2}'' (dolly \#100).
    
    \item \textsc{Storage Issue} reflects the data storage issues in memory encountered by practitioners of LLM open-source software, which primarily involve issues related to memory allocation strategies, anomalies or failures in data storage. For example, a user reported that ``\textit{memory allocation error occurred}'' when trying to ``\textit{inference on the given model}'' (unilm \#729).
\end{itemize}
\textbf{Interpretation}: \textit{Memory Issue} covers anomalies in the allocation and management of system memory during the operation of LLM open-source software. An LLM contains billions of parameters, and deploying and running LLM open-source software requires plenty of memory. \textit{Memory Issue} reflects the substantial demand for memory resources by LLM open-source software, which poses a challenge for practitioners with limited resources.
%\textsc{Out of Memory} (21) is the dominant type of \textit{Memory Issue}, which indicates the requirements of substantial memory resources for LLM open-source software to operate effectively.  %is the most frequent \textit{Memory Issue}. LLM open-source software has high memory resource requirements, which imposes significant demands on users' devices and leads to \textit{Memory Issue}.

\subsubsection{\textit{Prompt Issue (3.62\%)}} \textit{Prompt Issue} refers to issues concerning the prompts input into LLM open-source software and the process of inputting prompts. This category encompasses the following two types.
\begin{itemize}
    \item \textsc{Prompt Failure} refers to issues encountered when inputting prompts into LLM open-source software, including exceeding the length limit of prompts, using an invalid prompt data type, containing invalid characters in prompts, or errors when uploading prompts. For example, a user reported that ``\textit{the error of `valid\_languages' occurred}'' when uploading a PDF document as a prompt (haystack \#874).
    
    \item \textsc{Prompt Parsing Issue} refers to abnormalities and errors encountered by practitioners of LLM open-source software when parsing prompt content. For example, a user reported that the software ``\textit{only recognized the first four rows of a CSV file}'' when chatting with the model (langchain \#3621).
\end{itemize}
\textbf{Interpretation}: \textit{Prompt Issue} covers issues that arise when inputting prompts into LLM open-source software. \textsc{Prompt Failure} is the largest type (28), which reflects the restrictions of LLM open-source software regarding the file format and content of the prompts.
%Various LLM open-source software may have specific requirements for input prompts, such as limits on the number of tokens allowed in a prompt. If a prompt does not meet these requirements, it may lead to \textsc{Prompt Failure} (28).

\subsubsection{\textit{Security Issue (3.52\%)}} \textit{Security Issue} refers to security problems or vulnerabilities in LLM open-source software, including flaws, weaknesses, or errors within the system that could be exploited by malicious users to compromise its security. This category encompasses the following five types.
\begin{itemize}
    \item \textsc{User Permission Issue} refers to issues regarding user permissions of LLM open-source software. Many users have expressed concerns regarding user permissions, questioning whether user permissions are either insufficient or excessive. For example, a user had ``\textit{no right to download checkpoint}'', a file that stores the weights and state of the model (FastChat \#94).
    
    \item \textsc{Access Control Issue} refers to issues concerning the restriction of user access to data and system resources in LLM open-source software. The discussions on \textsc{Access Control Issue} aim to ensure the security and integrity of the software. For example, a user reported that ``\textit{it seems that GPT4ALL disallows}'' the interaction between the local webpage and the model, ``\textit{and the access is blocked by CORS}'' (gpt4all \#941).
    
    \item \textsc{Illegal Instruction} refers to the utilization of unauthorized instructions during the operation of LLM open-source software, which may potentially lead to security issues. For example, a user received a message that ``\textit{thread 5 `llm thread' received signal SIGILL, Illegal instruction}'', which shows that an illegal instruction made the core dump (gpt4all \#378).
    
    \item \textsc{Privacy Issue} refers to the risk of privacy breaches encountered by users when using LLM open-source software. For example, a user complained that ``\textit{different users share the same chat history}'' when chatting with the model, which shows the concern of users about their privacy (text-generation-webui \#2950).
    
    \item \textsc{Malware Issue} refers to the potential presence of malicious software attacks in LLM open-source software, which adversely affects the security of LLM open-source software. For example, a user reported that ``\textit{Windows defender claims to have found a backdoor in pytorch\_model.bin}'' (text-generation-webui \#456).
\end{itemize}
\textbf{Interpretation}: \textit{Security Issue} covers various security-related problems in LLM open-source software. As LLM open-source software becomes widespread, any developer can access and modify the source code or redeploy the LLM open-source software, which presents security challenges to practitioners, such as \textsc{User Permission Issue} (16), \textsc{Privacy Issue} (4), and \textsc{Malware Issue} (2). The proportion of \textit{Security Issues} in LLM open-source software is currently relatively small, but they cannot be overlooked, as these issues may lead to malicious exploitation of the software system, affecting the overall security of the software.

\subsubsection{\textit{Graphical User Interface (GUI) Issue (2.72\%)}} \textit{GUI Issue} refers to issues concerning the user interface of LLM open-source software. GUI serves as the channel through which users interact with the software and significantly impacts the user experience. This category encompasses the following two types.
\begin{itemize}
    \item \textsc{GUI Design Issue} refers to issues concerning the GUI design of LLM open-source software. A well-designed GUI enhances user experience, making optimization of GUI design an aspect worthy of attention. For example, a user suggested the development team ``\textit{activating extensions to appearing in GUI}'' that allow users selecting chat models more easily on their own (text-generation-webui \#2638).
    
    \item \textsc{GUI Failure} refers to the malfunction of the GUI in LLM open-source software, which emphasizes the maintenance required for the GUI. For example, a user reported that ``\textit{`File Upload' button not working}'' when chatting with the model (ChatDev \#227).
\end{itemize}
\textbf{Interpretation}: The design and maintenance of the GUI in LLM open-source software are closely tied to user experience, and consequently \textit{GUI Issue} is a concern for practitioners. \textsc{GUI Design Issue} (21) is the most common \textit{GUI Issue}. This type of issues reflects that the GUI design of LLM open-source software involves complex interactions that are not user-friendly and negatively impact the user experience. Developers may focus more on implementing functionality that uses LLMs and overlook the detailed design of the user interface. As a result, insufficient attention has been given to the testing and maintenance of the GUI, leading to \textit{GUI Issues}.

\subsubsection{\textit{Database Issue (2.41\%)}} \textit{Database Issue} refers to issues concerning the database of LLM open-source software. This category encompasses the following three types.
\begin{itemize}
    \item \textsc{Database Query Issue} refers to the issues concerning database queries in LLM open-source software. For example, a user reported that ``\textit{the table never changes, even if I change the custom\_query like changing the fields}'' when the user wanted to retrieve metadata for data preprocessing (haystack \#685).
    
    \item \textsc{Database Connection Issue} refers to the issues regarding the connection between the database and LLM open-source software, such as database connection interruptions. For example, a user reported that the connection to the database was lost when ``\textit{resetting the prompt template}'' (langchain \#2396).
    
    \item \textsc{Database Design Issue} refers to the issues concerning the design of database tables in LLM open-source software. For example, a user said that the number of fields in the ``\textit{onboarding}'' database table is insufficient to describe the status of the agents, and the table should at least include the following fields: ``\textit{user\_id, onboarding\_A, onboarding\_B\_1, onboarding\_B\_2, onboarding\_B\_3}'' (quivr \#1328).
\end{itemize}
\textbf{Interpretation}: \textit{Database Issue} covers various problems related to the databases in LLM open-source software, which can affect the data storage of LLM open-source software. \textsc{Database Query Issue} (16) is the dominant type of \textit{Database Issue}, which reflects the difficulties posed by the large-scale data within LLM open-source software, leading to inefficiencies in query execution and inaccuracies in query results.

\subsection{Category of Causes (RQ2)}\label{subsec:causesresults}

\subsubsection{\textbf{Results}} As mentioned in Section \ref{sec:formalDataExtraction}, not all closed issues have an associated cause that can be identified. As a result, we identified a total of 559 causes, which were extracted from 56.2\% of the 994 issues, and categorized into 11 categories as presented in Table \ref{Causes of Issues about LLM open-source projects}. The results show that the most prominent cause is \textit{Model Problem} (18.6\%), followed by \textit{Configuration and Connection Problem} (17.7\%), and \textit{Feature and Method Problem} (16.6\%). The examples of each category of cause with their counts and proportions are also provided in Table \ref{Causes of Issues about LLM open-source projects}. It should be noted that certain categories of issues can potentially be the causes of other issues. For example, \textsc{Model Collaboration Issue} is a type of \textit{Model Issue}, and it can also be the cause of other types of issue such as \textsc{Poor-quality Answer} and \textsc{Blank Output}. 

\begin{table*}[htbp]
\scriptsize
\caption{Causes of issues about LLM open-source projects}
\label{Causes of Issues about LLM open-source projects}
\begin{tabular}{m{5cm}m{9cm}m{0.8cm}<{\centering}m{0.8cm}<{\centering}}
\hline
\textbf{Category}                                           & \textbf{Example (Extracted Cause)} 
& \textbf{Count}   & \textbf{\%}           \\ \hline
Model Problem (MP)                                          & \textit{My first guess would be that you are running an ancient version of Transformers.} (text-generation-webui \#1650)    
& 104              & 18.6\% \\ \hline
Configuration and Connection Problem (CCP)                  & \textit{This looks like a fastchat issue. From the error above, there might be python environment issues.} (autogen \#207)        
& 99               & 17.7\% \\ \hline
Feature and Method Problem (FMP)                            & \textit{Providing validation can be extremely useful to users, especially less experienced ones, or users with very complex pipeline designs.} (haystack \#1981)                          
& 93               & 16.6\% \\ \hline
Parameter Problem (PP)                                      & \textit{The error message indicates that the `api\_url' argument is no longer accepted by the `anthropic.Client' initializer. } (langchain \#6883)     
& 72               & 12.9\% \\ \hline
Low Performance Problem (LPP)                               & \textit{Although the results converged after only a few epochs, it still is very slow.} (unilm \#144)   
& 42               & 7.5\% \\ \hline
Component Problem (CP)                                      & \textit{Is there a dedicated document loader tool in AutoGen, $\cdots$} (autogen \#607)   & 36               & 6.4\% \\ \hline
Incorrect Operation (IO)                                    & \textit{If you try to import the PDFToTextConverter without having the [ocr] option installed, it will give you an error.} (haystack \#4189)             
& 26               & 4.7\% \\ \hline
Hardware and Software Environment Incompatibility (HSEI)    & \textit{Looks like I'm missing the AVX instruction on my CPU.} (gpt4all \#378)    
& 22               & 3.9\% \\ \hline
Resource Problem (RP)                                       & \textit{In this case it's OS memory; I think 64GB is actually not enough.} (dolly \#194)     
& 20               & 3.6\% \\ \hline
Documentation Problem (DP)                                  & \textit{You must follow the three steps: $\cdots$} (FastChat \#186)
& 20               & 3.6\% \\ \hline
Security Problem (SP)                                       & \textit{Langchain agents can be hijacked while searching internet via injection prompts.} (langchain \#4102) 
& 15               & 2.7\% \\ \hline
\end{tabular}
\end{table*}

\begin{itemize}
    \item \textit{Model Problem} (MP) refers to problems concerning the models integrated into LLM open-source software. As the core component of LLM open-source software, the models are the most frequent source of issues. It encompasses five types: 
    (1) \textsc{Model Incompatibility}: Incompatibility between models, between models and other components, and between models and hardware. For example, a user received a message that ``\textit{Torch is not compatible with CUDA}'', leading to the failure of the installation of the LLM open-source software (text-generation-webui \#355).
    (2) \textsc{Model Training Problem}: Problems occurring when training models and problems with the training algorithms. 
    (3) \textsc{Outdated Model}: Using obsolete versions of models, and 
    (4) \textsc{Model Architecture Problem}: Optimization and refinement needed in model architecture implementations.

    \item \textit{Configuration and Connection Problem} (CCP) refers to problems concerning incorrect settings or failures in establishing connections within LLM open-source software. It encompasses three types: 
    (1) \textsc{Incorrect Configuration}: Failure of configuring LLM open-source software correctly. For example, a user failed to operate the localized LLMs, and found that ``\textit{there are python environment issues}'' after testing ``\textit{FastAPI}'', a framework for building Web APIs (autogen \#207).
    (2) \textsc{Network Connection Problem}: Issues arising from disruptions or failures in network connections of LLM open-source software, and 
    (3) \textsc{Database Connection Problem}: Problems in establishing connections with the database of LLM open-source software.

    \item \textit{Feature and Method Problem} (FMP) refers to problems concerning the functionality of LLM open-source software and its implementation methods. It encompasses two types: 
    (1) \textsc{Missing Feature}: The absence of one or more essential functions within LLM open-source software. For example, a user reported that the loss of a key feature of the retriever led to the issue that the model often ``\textit{gives non-sense answers}'' (haystack \#883), and 
    (2) \textsc{Incorrect Method Definition and Implementation}: Flaws in the logic, definition, and implementation of functions and methods of LLM open-source software.

    \item \textit{Parameter Problem} (PP) refers to problems concerning parameters within LLM open-source software. It encompasses two types: 
    (1) \textsc{Parameter Setting and Passing Problem}: Problems on the definition, assignment, and subsequent passing of parameters in LLM open-source software. For example, a user explained that ``\textit{the embedding\_function in model.encode() is returning}'' the array type not supported by the search engine, leading to the failure of the word embedding (langchain \#2016), and 
    (2) \textsc{Missing Parameter}: The absence of critical parameters in model training scripts, model hyperparameters, model configuration parameters, and inference scripts.

    \item \textit{Low Performance Problem} (LPP) refers to deficiencies in the performance of LLM open-source software. It encompasses five types: 
    (1) \textsc{Low Throughput}: Low data processing capability in LLM open-source software. For example, a user explained that ``\textit{Whisper API limits upload to 25MB}'', leading to the failure of preprocessing audio files (quivr \#3).
    (2) \textsc{Slow Response}: Response latency or timeout by LLM open-source software. 
    (3) \textsc{Low Scalability}: Poor scalability of LLM open-source software, limiting its ability to handle increasing loads. 
    (4) \textsc{Low Stability}: Runtime sensitivity of LLM open-source software to the software or hardware environment, and 
    (5) \textsc{Low Resource Utilization}: Inefficient resource utilization of LLM open-source software.

    \item \textit{Component Problem} (CP) refers to the problems among the various components of LLM open-source software. It can be categorized into three types: 
    (1) \textsc{Component Incompatibility}: Components are incompatible with other components within LLM open-source software. For example, the LangChain team explained that ``\textit{ZeroShotAgent does not support multi-input tool Calculator}'', which led to the failure of operating the model (langchain \#3781). 
    (2) \textsc{Missing Component}: The absence of crucial components within LLM open-source software, and 
    (3) \textsc{Outdated Component}: Outdated versions of components within LLM open-source software.

    \item \textit{Incorrect Operation} (IO) refers to the improper actions taken by practitioners of LLM open-source software. It can be categorized into three types: 
    (1) \textsc{Incorrect Operation Procedure}: Wrong steps in operating LLM open-source software. For example, the development team found that the user who raised the issue ``\textit{had skipped the llama to hugging face weight conversion step}'', leading to the failure of deploying the LLM open-source software (FastChat \#236). 
    (2) \textsc{Incorrect Method}: Utilizing incorrect functions for feature implementation of LLM open-source software, and 
    (3) \textsc{Incorrect Instruction}: Errors in instructions used to operate LLM open-source software.

    \item \textit{Hardware and Software Environment Incompatibility} (HSEI) refers to the problems encountered on the devices that LLM open-source software is deployed. It can be categorized into two types: 
    (1) \textsc{Hardware Incompatibility}: Incompatibility between hardware and LLM open-source software. For example, a user explained that ``\textit{I do not think 1080ti support 4int instruction set}'', leading to the failure of deployment of the LLM open-source software (gpt4all \#378), and 
    (2) \textsc{Software Environment Incompatibility}: Incompatibility between the software environment (e.g., operating system, virtual machine) and the LLM open-source software.

    \item \textit{Resource Problem} (RP) refers to the problems concerning the resources used by LLM open-source software. It can be categorized into two types: 
    (1) \textsc{Resource Shortage Problem}: Insufficient system resources of LLM open-source software. For example, a user reported that ``\textit{64GB is actually not enough}'' for training the model, and memory shortage led to the failure of model training (dolly \#194), and 
    (2) \textsc{Resource Allocation Problem}: Unreasonable system resource allocation of LLM open-source software.

    \item \textit{Documentation Problem} (DP) refers to the problems within the documentation of LLM open-source projects. It can be categorized into two types: 
    (1) \textsc{Poor Documentation}: The documentation of LLM open-source projects is incomplete, inaccurately described, lacking in detail, overly vague, or even nonexistent. For example, a user complained that ``\textit{add tutorials to increase token limits for Self-Depoly}'', as the loss of key documentation posed challenges for users in deploying the model (FastChat \#638), and 
    (2) \textsc{Outdated Documentation}: Documentation of LLM open-source projects becomes obsolete due to lack of updates or revisions.

    \item \textit{Security Problem} (SP) refers to problems concerning the security of LLM open-source software. It can be categorized into four types: 
    (1)\textsc{Access Problem}: Failure in security authentication when using LLM open-source software. For example, a user did not set an ``\textit{Elasticsearch client instance, index name, and embedding object}'' to obtain access permissions, resulting in the failure of environment configuration for the model (langchain \#1865). 
    (2) \textsc{Illegal Instruction}: Using unauthorized instructions to operate LLM open-source software. 
    (3) \textsc{User Permission Problem}: Insufficient or excessive permission for accessing and operating resources and functions in LLM open-source software, and 
    (4) \textsc{Malware Problem}: Attacks on LLM open-source software by malware.
\end{itemize}

\begin{table*}[h]
\scriptsize
\caption{Mapping between issue categories (vertical) and cause categories (horizontal)}
\label{tab:issues-causes}
\begin{adjustbox}{center}
\begin{tabular}{ccccccccccccc}
                                                    & \multicolumn{1}{l}{\textbf{MP}} & \multicolumn{1}{l}{\textbf{CCP}} & \multicolumn{1}{l}{\textbf{FMP}} 
                                                    & \multicolumn{1}{l}{\textbf{PP}} & \multicolumn{1}{l}{\textbf{LPP}} & \multicolumn{1}{l}{\textbf{CP}} 
                                                    & \multicolumn{1}{l}{\textbf{IO}} & \multicolumn{1}{l}{\textbf{HSEI}} & \multicolumn{1}{l}{\textbf{RP}} 
                                                    & \multicolumn{1}{l}{\textbf{DP}} & \multicolumn{1}{l}{\textbf{SP}} \\
\multirow{1}{*}{\cellcolor[HTML]{E0F7E0}\textbf{{Model Issue}}} 
                                                    & \cellcolor[HTML]{F2F2F2}0      & \cellcolor[HTML]{4472C4}25         & \cellcolor[HTML]{4472C4}21 
                                                    & \cellcolor[HTML]{4472C4}22      & \cellcolor[HTML]{8EAADB}15          & \cellcolor[HTML]{D9E2F3}10 
                                                    & \cellcolor[HTML]{D9E2F3}2       & \cellcolor[HTML]{D9E2F3}8         & \cellcolor[HTML]{D9E2F3}10 
                                                    & \cellcolor[HTML]{D9E2F3}6       & \cellcolor[HTML]{D9E2F3}4       \\
\multirow{1}{*}{\cellcolor[HTML]{FFFFE0}\textbf{{Component Issue}}}  
                                                    & \cellcolor[HTML]{8EAADB}15       & \cellcolor[HTML]{D9E2F3}5         & \cellcolor[HTML]{D9E2F3}2 
                                                    & \cellcolor[HTML]{8EAADB}11       & \cellcolor[HTML]{D9E2F3}3         & \cellcolor[HTML]{F2F2F2}0 
                                                    & \cellcolor[HTML]{D9E2F3}4       & \cellcolor[HTML]{D9E2F3}2         & \cellcolor[HTML]{F2F2F2}0 
                                                    & \cellcolor[HTML]{D9E2F3}3       & \cellcolor[HTML]{D9E2F3}3       \\
\multirow{1}{*}{\cellcolor[HTML]{E0F7E0}\textbf{{Parameter Issue}}} 
                                                    & \cellcolor[HTML]{4472C4}24      & \cellcolor[HTML]{8EAADB}13         & \cellcolor[HTML]{8EAADB}16 
                                                    & \cellcolor[HTML]{F2F2F2}0      & \cellcolor[HTML]{D9E2F3}5         & \cellcolor[HTML]{D9E2F3}6 
                                                    & \cellcolor[HTML]{D9E2F3}3       & \cellcolor[HTML]{D9E2F3}1         & \cellcolor[HTML]{F2F2F2}0 
                                                    & \cellcolor[HTML]{D9E2F3}3       & \cellcolor[HTML]{D9E2F3}1       \\
\multirow{1}{*}{\cellcolor[HTML]{FFFFE0}\textbf{{Answer Issue}}}  
                                                    & \cellcolor[HTML]{D9E2F3}9       & \cellcolor[HTML]{D9E2F3}9         & \cellcolor[HTML]{D9E2F3}4 
                                                    & \cellcolor[HTML]{D9E2F3}8       & \cellcolor[HTML]{D9E2F3}5         & \cellcolor[HTML]{D9E2F3}1 
                                                    & \cellcolor[HTML]{D9E2F3}5       & \cellcolor[HTML]{D9E2F3}1         & \cellcolor[HTML]{F2F2F2}0 
                                                    & \cellcolor[HTML]{D9E2F3}1       & \cellcolor[HTML]{D9E2F3}1       \\
\multirow{1}{*}{\cellcolor[HTML]{E0F7E0}\textbf{{Performance Issue}}}  
                                                    & \cellcolor[HTML]{8EAADB}14       & \cellcolor[HTML]{D9E2F3}2         & \cellcolor[HTML]{8EAADB}15 
                                                    & \cellcolor[HTML]{D9E2F3}3      & \cellcolor[HTML]{F2F2F2}0        & \cellcolor[HTML]{D9E2F3}2 
                                                    & \cellcolor[HTML]{D9E2F3}1       & \cellcolor[HTML]{D9E2F3}1         & \cellcolor[HTML]{D9E2F3}2 
                                                    & \cellcolor[HTML]{F2F2F2}0       & \cellcolor[HTML]{D9E2F3}2       \\
\multirow{1}{*}{\cellcolor[HTML]{FFFFE0}\textbf{{Code Issue}}} 
                                                    & \cellcolor[HTML]{D9E2F3}8       & \cellcolor[HTML]{D9E2F3}6         & \cellcolor[HTML]{8EAADB}11
                                                    & \cellcolor[HTML]{D9E2F3}8      & \cellcolor[HTML]{D9E2F3}2         & \cellcolor[HTML]{D9E2F3}3 
                                                    & \cellcolor[HTML]{D9E2F3}1       & \cellcolor[HTML]{F2F2F2}0         & \cellcolor[HTML]{D9E2F3}3 
                                                    & \cellcolor[HTML]{D9E2F3}1       & \cellcolor[HTML]{D9E2F3}1       \\
\multirow{1}{*}{\cellcolor[HTML]{E0F7E0}\textbf{{Installation Issue}}} 
                                                    & \cellcolor[HTML]{D9E2F3}6       & \cellcolor[HTML]{8EAADB}14        & \cellcolor[HTML]{F2F2F2}0 
                                                    & \cellcolor[HTML]{F2F2F2}0       & \cellcolor[HTML]{F2F2F2}0         & \cellcolor[HTML]{D9E2F3}5 
                                                    & \cellcolor[HTML]{D9E2F3}2       & \cellcolor[HTML]{D9E2F3}1         & \cellcolor[HTML]{F2F2F2}0 
                                                    & \cellcolor[HTML]{D9E2F3}1       & \cellcolor[HTML]{D9E2F3}1       \\
\multirow{1}{*}{\cellcolor[HTML]{FFFFE0}\textbf{{Documentation Issue}}} 
                                                    & \cellcolor[HTML]{F2F2F2}0       & \cellcolor[HTML]{F2F2F2}0         & \cellcolor[HTML]{D9E2F3}3 
                                                    & \cellcolor[HTML]{D9E2F3}1       & \cellcolor[HTML]{D9E2F3}1         & \cellcolor[HTML]{D9E2F3}1 
                                                    & \cellcolor[HTML]{F2F2F2}0       & \cellcolor[HTML]{F2F2F2}0         & \cellcolor[HTML]{F2F2F2}0 
                                                    & \cellcolor[HTML]{F2F2F2}0       & \cellcolor[HTML]{F2F2F2}0       \\
\multirow{1}{*}{\cellcolor[HTML]{E0F7E0}\textbf{{Configuration Issue}}} 
                                                    & \cellcolor[HTML]{D9E2F3}5       & \cellcolor[HTML]{D9E2F3}8         & \cellcolor[HTML]{D9E2F3}3 
                                                    & \cellcolor[HTML]{D9E2F3}4       & \cellcolor[HTML]{F2F2F2}0         & \cellcolor[HTML]{D9E2F3}2 
                                                    & \cellcolor[HTML]{D9E2F3}1       & \cellcolor[HTML]{F2F2F2}0         & \cellcolor[HTML]{D9E2F3}1 
                                                    & \cellcolor[HTML]{D9E2F3}3       & \cellcolor[HTML]{F2F2F2}0       \\
\multirow{1}{*}{\cellcolor[HTML]{FFFFE0}\textbf{{Network Issue}}} 
                                                    & \cellcolor[HTML]{D9E2F3}4      & \cellcolor[HTML]{D9E2F3}7          & \cellcolor[HTML]{D9E2F3}4 
                                                    & \cellcolor[HTML]{D9E2F3}5      & \cellcolor[HTML]{D9E2F3}1          & \cellcolor[HTML]{D9E2F3}1 
                                                    & \cellcolor[HTML]{F2F2F2}0      & \cellcolor[HTML]{D9E2F3}2          & \cellcolor[HTML]{F2F2F2}0 
                                                    & \cellcolor[HTML]{F2F2F2}0      & \cellcolor[HTML]{D9E2F3}2       \\
\multirow{1}{*}{\cellcolor[HTML]{E0F7E0}\textbf{{Memory Issue}}} 
                                                    & \cellcolor[HTML]{D9E2F3}9      & \cellcolor[HTML]{D9E2F3}3          & \cellcolor[HTML]{D9E2F3}4 
                                                    & \cellcolor[HTML]{D9E2F3}3      & \cellcolor[HTML]{D9E2F3}1          & \cellcolor[HTML]{D9E2F3}1
                                                    & \cellcolor[HTML]{F2F2F2}0      & \cellcolor[HTML]{D9E2F3}1         & \cellcolor[HTML]{D9E2F3}4 
                                                    & \cellcolor[HTML]{F2F2F2}0      & \cellcolor[HTML]{F2F2F2}0       \\
\multirow{1}{*}{\cellcolor[HTML]{FFFFE0}\textbf{{Prompt Issue}}} 
                                                    & \cellcolor[HTML]{D9E2F3}4      & \cellcolor[HTML]{D9E2F3}3          & \cellcolor[HTML]{D9E2F3}2 
                                                    & \cellcolor[HTML]{D9E2F3}2      & \cellcolor[HTML]{D9E2F3}5          & \cellcolor[HTML]{D9E2F3}2 
                                                    & \cellcolor[HTML]{D9E2F3}2      & \cellcolor[HTML]{D9E2F3}1          & \cellcolor[HTML]{F2F2F2}0 
                                                    & \cellcolor[HTML]{D9E2F3}2      & \cellcolor[HTML]{F2F2F2}0       \\
\multirow{1}{*}{\cellcolor[HTML]{E0F7E0}\textbf{{Security Issue}}} 
                                                    & \cellcolor[HTML]{D9E2F3}2      & \cellcolor[HTML]{D9E2F3}1          & \cellcolor[HTML]{D9E2F3}3 
                                                    & \cellcolor[HTML]{D9E2F3}3      & \cellcolor[HTML]{D9E2F3}1          & \cellcolor[HTML]{D9E2F3}2 
                                                    & \cellcolor[HTML]{D9E2F3}3      & \cellcolor[HTML]{D9E2F3}3          & \cellcolor[HTML]{F2F2F2}0 
                                                    & \cellcolor[HTML]{F2F2F2}0      & \cellcolor[HTML]{F2F2F2}0       \\
\multirow{1}{*}{\cellcolor[HTML]{FFFFE0}\textbf{{GUI Issue}}} 
                                                    & \cellcolor[HTML]{F2F2F2}0      & \cellcolor[HTML]{D9E2F3}3          & \cellcolor[HTML]{D9E2F3}2 
                                                    & \cellcolor[HTML]{F2F2F2}0      & \cellcolor[HTML]{F2F2F2}0          & \cellcolor[HTML]{F2F2F2}0 
                                                    & \cellcolor[HTML]{D9E2F3}1      & \cellcolor[HTML]{D9E2F3}1          & \cellcolor[HTML]{F2F2F2}0 
                                                    & \cellcolor[HTML]{F2F2F2}0      & \cellcolor[HTML]{F2F2F2}0       \\
\multirow{1}{*}{\cellcolor[HTML]{E0F7E0}\textbf{{Database Issue}}} 
                                                    & \cellcolor[HTML]{D9E2F3}4      & \cellcolor[HTML]{F2F2F2}0          & \cellcolor[HTML]{D9E2F3}3 
                                                    & \cellcolor[HTML]{D9E2F3}2      & \cellcolor[HTML]{D9E2F3}3          & \cellcolor[HTML]{F2F2F2}0 
                                                    & \cellcolor[HTML]{D9E2F3}1      & \cellcolor[HTML]{F2F2F2}0          & \cellcolor[HTML]{F2F2F2}0 
                                                    & \cellcolor[HTML]{F2F2F2}0      & \cellcolor[HTML]{F2F2F2}0       \\
\end{tabular}
\end{adjustbox}
\begin{minipage}{18cm} 
\vspace{0.1cm}
\vspace{0.1cm}
  \textbf{Full names of every cause category abbreviation}: MP: \textit{Model Problem}; CCP: \textit{Configuration and Connection Problem}; PP: \textit{Parameter Problem}; FMP: \textit{Feature and Method Problem}; LPP: \textit{Low Performance Problem}; CP: \textit{Component Problem}; IO: \textit{Incorrect Operation}; HSEI: \textit{Hardware and Software Environment Incompatibility}; RP: \textit{Resource Problem}; DP: \textit{Documentation Problem}; SP: \textit{Security Problem}.
\end{minipage}
\end{table*}

\subsubsection{Causes to Issues Mapping}
Table~\ref{tab:issues-causes} illustrates the mapping between the \textit{Issue} categories and \textit{Cause} categories in developing and using LLM open-source software, along with their distribution, using abbreviations to represent each category of \textit{Cause}. For example, ``MP'' represents \textit{Model Problem}. The full names of all cause categories are provided in the note of Table \ref{tab:issues-causes}.

Over three quarters (78.3\%) of \textit{Parameter Issues} are identified with associated causes, reaching the highest rate among all categories of issues, and \textit{Parameter Issues} are mainly caused by MP (24).

Only 12.8\% of \textit{Documentation Issue} are identified with corresponding causes, having the lowest rate among all categories of issues. There are only a few cases of \textit{Documentation Issues} have their causes, and FMP (3) is identified as the most frequent cause.

The largest category among all the issues is \textit{Model Issue}. 50.4\% of \textit{Model Issues} are identified with the corresponding causes. \textit{Model Issue} mainly originates from CCP (25), FMP (21), and PP (22). 

For \textit{Component Issue}, causes are identified in 52.2\% of cases. \textit{Component Issue} is mainly caused by MP (15).

For \textit{Answer Issue}, causes are identified in 53.7\% of cases. MP (9) and CCP (9) are main causes leading to \textit{Answer Issue}.

For \textit{Performance Issue}, causes are identified in 59.2\% of cases. FMP (15) is identified as the most frequent cause of \textit{Performance Issue}.

For \textit{Code Issue}, causes are identified in 65.7\% of cases. \textit{Code Issue} is mainly induced by FMP (11).

For \textit{Installation Issue}, causes are identified in 61.2\% of cases. \textit{Installation Issue} mainly originates from CCP (14).

For \textit{Configuration Issue}, causes are identified in 60\% of cases. \textit{Configuration Issue} is mainly caused by CCP (8).

For \textit{Network Issue}, causes are identified in 59.1\% of cases. \textit{Network Issue} is mainly brought about by CCP (7).

For \textit{Memory Issue}, causes are identified in 66.7\% of cases. \textit{Memory Issue} is mainly induced by MP (9).

For \textit{Prompt Issue}, causes are identified in 63.9\% of cases. LPP (5) is identified as the most frequent cause of \textit{Prompt Issue}. 

For \textit{Security Issue}, causes are identified in 51.4\% of cases. \textit{Security Issue} is mainly caused by FMP (3), PP (3), IO (3), and HSEI (3).

For \textit{GUI Issue}, causes are identified in 25.9\% of cases. \textit{GUI Issue} mainly originates from CCP (3).

For \textit{Database Issue}, causes are identified in 54.2\% of cases. \textit{Database Issue} is mainly brought about by MP (4).

\subsubsection{\textbf{Interpretation}} 
\textbf{Frequency of the \textit{causes}}: MP is the most frequent factor contributing to issues. As core functional components of LLM open-source software, the fact that MP is the most frequent cause of issues shows that the challenges faced by the models remain formidable. \textit{Model Issue} stems from various factors such as model architecture, updates, training, and resource requirements, making the use and maintenance of LLMs challenging for practitioners.
In addition, CCP and FMP also significantly contribute to the issues. These types of causes are general problems in the development of LLM open-source software. 
The process of integrating LLMs into software to implement functionality is highly complex. When selecting components, users may encounter incompatibility between components, between components and models, and between components and the software or hardware environment.  
In the implementation of functionalities, design flaws of functions and classes can arise. In addition, LLM open-source software contains complex algorithms, which presents challenges in implementing the features of LLM open-source software.
Although each of the remaining eight categories of causes of the issues accounts for a relatively minor proportion, they also cannot be ignored. For example, during the definition and settings of parameters, there may be problems related to incorrect parameter settings, such as improper values, improper data type, missing parameters, and redundant parameters. PP reflects that the large-scale parameters of LLMs introduce difficulties in the development and maintenance of LLM open-source software.

\textbf{Mapping of \textit{causes} to \textit{issues}}: Nearly a quarter (24.0\%) of \textit{Model Issues} are caused by CCP. Model version updates and differences in model types may lead to incompatibility between the components and models. When using different models or frameworks, or after model updates, parameter formats or inference mechanisms may be changed, which can also results in incompatibility with other components.
20.2\% of \textit{Model Issues} are induced by FMP, which reflects the complexity of the design and implementation of features of LLM open-source software. The accuracy of responses, the performance of models, and execution of training or inference in LLMs rely on the proper implementation of functionalities. If the functional design is flawed or the implementation method is incorrect, the performance of models will be adversely affected.
21.2\% of \textit{Model Issues} originate from PP, and 33.3\% of \textit{Parameter Issues} are caused by MP. The design and execution of the model rely on parameter settings, if the parameters are not set properly, the model may fail to function correctly. LLMs, with their large-scale parameters, make parameter settings more difficult, and the design flaws and limitations of models (such as restrictions on parameter length) bring specific constraints to parameter settings, leading to \textit{Parameter Issues}.

\subsection{Category of Solutions (RQ3)}
\label{subsec:solutionsresults}

\subsubsection{\textbf{Results}}
As mentioned in Section \ref{sec:formalDataExtraction}, not all closed issues have an associated solution that can be identified. As a result, we identified a total of 798 solutions, which were used to address 80.3\% of the 994 issues, and categorized into seven categories as presented in Table \ref{Solutions of Issues about LLM open-source projects}. The results show that \textit{Optimize Model} (OM) is the most commonly utilized solution, accounting for 29.6\% of the total solutions; closely followed by \textit{Adjust Parameter} (AP), \textit{Adjust Configuration and Operation} (ACO), and \textit{Optimize Component} (OC), accounting for 19.4\%, 18.0\%, and 17.2\% respectively. The examples of each category of solution with their counts and proportions are provided in Table \ref{Solutions of Issues about LLM open-source projects}. 

\begin{itemize}
    \item \textit{Optimize Model} (OM) refers to the process of solving problems by fine-tuning or refining the LLMs. It is also the most widely accepted solution for addressing problems. It can be categorized into five types: 
    (1) \textsc{Optimize Model Architecture}: Improving the architectural design and implementation of LLMs. For example, a user suggested that ``\textit{it is better to convert that model to llama CPP ggml and use it with the wrapper}'' to address the issue of the inability of the model to operate on Intel Core i7 CPU. (text-generation-webui \#1233). 
    (2) \textsc{Optimize Model Training}: Enhancing the training methods of LLMs, such as improving training algorithms and adjusting relevant process controls when training models. 
    (3) \textsc{Update Model}: Updating the models by replacing old versions with newer ones. 
    (4) \textsc{Use Stable Version of Model}: Utilizing a version of the LLM that offers stable functionality and performance to ensure stability, and 
    (5) \textsc{Optimize Dataset}: Selecting datasets, which are more widely used by practitioners, for training, fine-tuning, and evaluating the models integrated into LLM open-source software.
    
    \item \textit{Adjust Parameter} (AP) refers to resolving issues by adjusting parameters. Parameters play a crucial role in software systems, especially in LLM open-source software, which involves billions of parameters. Therefore, it accounts for a significant proportion. It can be categorized into five types: 
    (1) \textsc{Modify Parameter}: Modifying the definition and values of parameters within LLM open-source software to address the problem. For example, a user suggested ``\textit{setting `infer\_tokenizer\_classes = True' when initializing DPR (Dense Passage Retrieval)}'' to address the issue of freezing during the preprocessing of the training dataset (haystack \#822). 
    (2) \textsc{Remove Parameter}: Removing redundant parameters within LLM open-source software. 
    (3) \textsc{Add Parameter}: Adding parameters to fill in missing parameters or adding parameters to control new functionality for LLM open-source software. 
    (4) \textsc{Convert Parameter Type}: Converting parameter types during parameter passing to match subsequent parameter processing of LLM open-source software, and 
    (5) \textsc{Add Parameter Validation}: Adding parameter validation to ensure the correctness of functionalities within LLM open-source software. 
    
    \item \textit{Adjust Configuration and Operation} (ACO) refers to resolving issues by modifying software configurations and user operations. Proper configuration operation are crucial for the normal functioning of LLM open-source software. It can be categorized into four types: 
    (1) \textsc{Modify Incorrect Operation Procedure}: Adjusting operation procedures by following related documentation and tutorials to operate LLM open-source software. For example, the development team suggested users ``\textit{initializing an empty tool list first, then adding a custom tool}'' to address the custom tool failure (langchain \#2569). 
    (2) \textsc{Modify Configuration}: Adjusting configurations of LLM open-source software, including environmental settings, initialization configurations. 
    (3) \textsc{Reload Project}: Reloading or restarting LLM open-source software, and 
    (4) \textsc{Modify Incorrect Instruction}: Correct errors in instructions for operating LLM open-source software.
    
    \item \textit{Optimize Component} (OC) refers to the adjustment and enhancement of various components within LLM open-source software. It can be categorized into six types: 
    (1) \textsc{Add New Component}: Introducing new components to implement new functionalities for LLM open-source software. For example, a user announced that ``\textit{I created the `chatmemorysources' container}'' to address the failure of uploading the documents when chatting with the model (semantic-kernel \#1402). 
    (2) \textsc{Adjust Component Integration Strategy}: Modifying the strategies used to integrate various components within LLM open-source software, such as component decoupling and adjusting interfaces between components, to enhance the stability, scalability, and performance of LLM open-source software. 
    (3) \textsc{Optimize Component Functionality}: Optimizing specific component functionalities within LLM open-source software to address corresponding issues. 
    (4) \textsc{Use Stable Version of Component}: Using stable versions of components to ensure relatively stable functionality and performance of LLM open-source software. 
    (5) \textsc{Update Component}: Incorporating newer versions of components to iterate functionalities of LLM open-source software, and 
    (6) \textsc{Remove Component}: Eliminating redundant, outdated, inefficient, incompatible, and vulnerable components of LLM open-source software.
    
    \item \textit{Optimize System Resource Management} (OSRM) refers to adjusting the strategies of managing system resources for operating LLM open-source software. It can be categorized into four types: 
    (1) \textsc{Optimize Network Connection Management}: Adjusting network connection settings and protocol implementations to ensure that LLM open-source software runs in an optimal network environment. For example, a user suggested that ``\textit{you can stop the generation by posting to the server}'' instead of using Server-Sent Events (SSE) to address the issue that ``\textit{closing http socket does not cancel generation}'' (text-generation-webui \#4521). 
    (2) \textsc{Optimize Memory Management}: Adjusting memory allocation and management strategies for LLM open-source software. 
    (3) \textsc{Optimize Message Management}: Adjusting the messaging management mechanism within LLM open-source software, including aspects such as pipe communication and message passing, and
    (4) \textsc{Optimize Process Management}: Optimizing process management (e.g., concurrent processing) and resource allocation strategies within LLM open-source software to reduce waiting time, minimize resource waste, and prevent deadlocks.
    
    \item \textit{Modify Documentation} (MD) refers to altering and refining the documentation of LLM open-source projects. It can be categorized into three types: 
    (1) \textsc{Modify Description}: Modifying descriptions in the documentation of LLM open-source projects, involving correcting wrong, vague, outdated, and inconsistent descriptions within the documentation. For example, a user suggested ``\textit{separating all commands and text to improve readability}'', which can address the issue of unclear descriptions in the deployment tutorial of the LLM open-source project (haystack \#1863). 
    (2) \textsc{Add Documentation}: Adding new documentation of LLM open-source projects to synchronize documentation with the software updates, and 
    (3) \textsc{Rename Documents}: Modifying document names of LLM open-source projects that are ambiguous or contain illegal characters.
    
    \item \textit{Change Hardware and Software Environment} (CHSE) refers to modifying and optimizing the hardware and software environment for running operating LLM open-source software. It can be categorized into two types: 
    (1) \textsc{Change Hardware}: Replacing hardware for configuring LLM open-source projects, such as CPU, GPU, or even the entire computer. For example, the haystack team announced that ``\textit{we opt by changing the machine we are using to one with an NVIDIA Graphics Card}'' to address the failure of deploying the LLM open-source software (haystack \#1643), and 
    (2) \textsc{Change Software Environment}: Replacing software environment for operating LLM open-source projects, such as operating system, virtual machine, and container.
\end{itemize}

\begin{table*}[htbp]
\scriptsize
\caption{Solutions of issues about LLM open-source projects}
\label{Solutions of Issues about LLM open-source projects}
%\begin{tabular}{m{3.35cm}m{11.9cm}m{0.6cm}<{\centering}m{0.6cm}<{\centering}}
\begin{tabular}{m{5.4cm}m{8.6cm}m{0.8cm}<{\centering}m{0.8cm}<{\centering}}
\hline
\textbf{Category}                              & \textbf{Example (Extracted Solution)}                                                                      
& \textbf{Count}            & \textbf{\%}           \\ \hline
Optimize Model (OM)                        & \textit{I think its better to convert that model to llama CPP ggml and use it on CPU that way with the wrapper.} (text-generation-webui \#1233)                                                   
& 236                       & 29.6\% \\ \hline
Adjust Parameter (AP)                      & \textit{To fix this issue, you need to replace `api\_url' with `base\_url' in the `Anthropic' class in your LangChain code.} (langchain \#6883)        
& 155                       & 19.4\% \\ \hline
Adjust Configuration and Operation (ACO)   & \textit{Have you tried cloning the model repository into your models directory and loading it that way?} (text-generation-webui \#3656)     
& 144                       & 18.0\% \\ \hline
Optimize Component (OC)                    & \textit{We will also soon release a high throughput batching backend.} (FastChat \#1041)   
& 137                       & 17.2\% \\ \hline
Optimize System Resource Management (OSRM) & \textit{The workaround currently is to manually increase the swap space on your drives to be at least 140GB, but you can set a minimum to like 16mb so its not always huge.} (text-generation-webui \#1647)             
& 63                        & 7.9\% \\ \hline
Modify Documentation (MD)                  & \textit{In addition, let's make sure that the use case of DocumentRetrieval pipelines as mentioned in \#1618 is properly documented.} haystack \#1863)                          
& 54                        & 6.8\% \\ \hline
Change Hardware and Software Environment (CHSE)                         & \textit{We will opt by changing the machine we're using to one with an NVIDIA Graphics Card.} (haystack \#1643)  
& 9                         & 1.1\% \\ \hline
\end{tabular}
\end{table*}

\subsubsection{Solutions to Issues Mapping}

Table~\ref{tab:issues-solutions} illustrates the mapping of LLM open-source projects related \textit{Issue} categories to \textit{Solution} categories, along with their distribution, using abbreviations to represent each category of solution. For example, ``OM'' represents \textit{Optimize Model}. The full names of all solution categories are provided in the note of Table \ref{tab:issues-solutions}.

%In all categories of issues, the highest resolution rate is for \textit{Code Issue}, reaching 92.5\%. OM is the solution with the largest proportion (40.3\%), followed by AP (22.6\%).
For \textit{Code Issue}, 92.5\% of cases are well addressed, reaching the highest resolution rate among all categories of issues. OM (25) is identified as the most frequent solution for resolving \textit{Code Issues}.

For \textit{Component Issue}, 70.7\% of the cases are effectively addressed, reaching the lowest resolution rate among all categories of issues, and \textit{Component Issues} are mostly solved by OC (27).
%In all categories of issues, the lowest resolution rate is found in \textit{Component Issue}, with only 70.7\% resolved. The most frequently used solution of \textit{Component Issue} is OC (41.5\%). 

For \textit{Model Issue}, which is the largest category among all categories of issues, 80.1\% of cases have their effective solutions. \textit{Model Issues} are mostly addressed by OM (88). 
%The largest category among all the issues is \textit{Model Issue}, with a resolution rate of 80.1\%. The most commonly used solution is OM (45.6\%). Other solutions, such as ACO addressing issues through configuration and operation (19.2\%), and OC tackling problems from a component perspective (12.4\%), are also widely accepted by practitioners.

For \textit{Parameter Issue}, 83.7\% of the cases are well addressed. The most frequently used solution for \textit{Parameter Issues} is AP (36).

For \textit{Answer Issue}, 86.6\% of cases are identified with their solutions. OM (20) is mostly employed solution to resolve \textit{Answer Issues}.

For \textit{Performance Issue}, 77.5\% cases are well resolved. The most frequently employed solution for \textit{Performance Issues} is OM (21).
 
For \textit{Installation Issue}, 81.6\% of cases have their effective solutions. ACO (17) is identified as the most frequently used solution to address \textit{Installation Issues}. 

For \textit{Documentation Issue}, 74.5\% of the cases are identified with corresponding solutions. The most frequently employed solution for \textit{Documentation Issues} is MD (23). 

For \textit{Configuration Issue}, 75.6\% of cases are effectively addressed. The most frequently employed solution to address \textit{Configuration Issues} is ACO (10). 

For \textit{Network Issue}, 75\% of cases are addressed by practitioners. The most frequently employed solution for \textit{Network Issues} is AP (13).

For \textit{Memory Issue}, 89.7\% of cases are identified with their associated solutions. OM (11) is the most commonly applied solution to resolve \textit{Memory Issues}.

For \textit{Prompt Issue}, 77.8\% of cases are resolved. The most frequently used solution to address \textit{Prompt Issues} is OM (16). 

For \textit{Security Issue}, 77.1\% of cases are resolved by practitioners. ACO is identified as the most commonly applied solution to (9) address \textit{Security Issues}. 

For \textit{GUI Issue}, 77.8\% of cases are well addressed by practitioners. The most commonly employed solution for \textit{GUI Issues} is OC (14).

For \textit{Database Issue}, 91.7\% of cases are well resolved by practitioners. OC is identified as the most frequently employed solution for \textit{Database Issues}.

\begin{table*}[h]
\scriptsize
\caption{ Mapping between issue categories (vertical) and solution categories (horizontal)}
\label{tab:issues-solutions}
\begin{adjustbox}{center}
\begin{tabular}{ccccccccc}
                                          & \multicolumn{1}{l}{\textbf{OM}} & \multicolumn{1}{l}{\textbf{AP}}   & \multicolumn{1}{l}{\textbf{ACO}} 
                                          & \multicolumn{1}{l}{\textbf{OC}} & \multicolumn{1}{l}{\textbf{OSRM}} & \multicolumn{1}{l}{\textbf{MD}} 
                                          & \multicolumn{1}{l}{\textbf{CHSE}} &  \\
\multirow{1}{*}{\cellcolor[HTML]{E0F7E0}\textbf{{Model Issue}}} 
                                          & \cellcolor[HTML]{4472C4}88      & \cellcolor[HTML]{8EAADB}20        & \cellcolor[HTML]{4472C4}37 
                                          & \cellcolor[HTML]{4472C4}24      & \cellcolor[HTML]{8EAADB}13        & \cellcolor[HTML]{D9E2F3}7 
                                          & \cellcolor[HTML]{D9E2F3}4         \\
\multirow{1}{*}{\cellcolor[HTML]{FFFFE0}\textbf{{Component Issue}}}  
                                          & \cellcolor[HTML]{D9E2F3}8       & \cellcolor[HTML]{8EAADB}14        & \cellcolor[HTML]{8EAADB}12 
                                          & \cellcolor[HTML]{4472C4}27      & \cellcolor[HTML]{D9E2F3}1         & \cellcolor[HTML]{D9E2F3}3 
                                          & \cellcolor[HTML]{F2F2F2}0          \\
\multirow{1}{*}{\cellcolor[HTML]{E0F7E0}\textbf{{Parameter Issue}}} 
                                          & \cellcolor[HTML]{8EAADB}15      & \cellcolor[HTML]{4472C4}36        & \cellcolor[HTML]{8EAADB}13 
                                          & \cellcolor[HTML]{D9E2F3}4       & \cellcolor[HTML]{D9E2F3}5         & \cellcolor[HTML]{D9E2F3}4 
                                          & \cellcolor[HTML]{F2F2F2}0            \\
\multirow{1}{*}{\cellcolor[HTML]{FFFFE0}\textbf{{Answer Issue}}}  
                                          & \cellcolor[HTML]{8EAADB}20      & \cellcolor[HTML]{8EAADB}18        & \cellcolor[HTML]{8EAADB}16 
                                          & \cellcolor[HTML]{8EAADB}11      & \cellcolor[HTML]{D9E2F3}5         & \cellcolor[HTML]{D9E2F3}1 
                                          & \cellcolor[HTML]{F2F2F2}0          \\
\multirow{1}{*}{\cellcolor[HTML]{E0F7E0}\textbf{{Performance Issue}}}  
                                          & \cellcolor[HTML]{4472C4}21      & \cellcolor[HTML]{8EAADB}11        & \cellcolor[HTML]{D9E2F3}3 
                                          & \cellcolor[HTML]{8EAADB}12      & \cellcolor[HTML]{D9E2F3}7         & \cellcolor[HTML]{D9E2F3}1 
                                          & \cellcolor[HTML]{F2F2F2}0         \\
\multirow{1}{*}{\cellcolor[HTML]{FFFFE0}\textbf{{Code Issue}}} 
                                          & \cellcolor[HTML]{4472C4}25      & \cellcolor[HTML]{8EAADB}14        & \cellcolor[HTML]{D9E2F3}8 
                                          & \cellcolor[HTML]{D9E2F3}6       & \cellcolor[HTML]{D9E2F3}6         & \cellcolor[HTML]{D9E2F3}3 
                                          & \cellcolor[HTML]{F2F2F2}0        \\
\multirow{1}{*}{\cellcolor[HTML]{E0F7E0}\textbf{{Installation Issue}}} 
                                          & \cellcolor[HTML]{D9E2F3}10      & \cellcolor[HTML]{D9E2F3}3         & \cellcolor[HTML]{8EAADB}17 
                                          & \cellcolor[HTML]{D9E2F3}6       & \cellcolor[HTML]{D9E2F3}1         & \cellcolor[HTML]{D9E2F3}2 
                                          & \cellcolor[HTML]{D9E2F3}1         \\
\multirow{1}{*}{\cellcolor[HTML]{FFFFE0}\textbf{{Documentation Issue}}} 
                                          & \cellcolor[HTML]{D9E2F3}2       & \cellcolor[HTML]{D9E2F3}2         & \cellcolor[HTML]{D9E2F3}5 
                                          & \cellcolor[HTML]{D9E2F3}2       & \cellcolor[HTML]{D9E2F3}1         & \cellcolor[HTML]{4472C4}23 
                                          & \cellcolor[HTML]{F2F2F2}0         \\
\multirow{1}{*}{\cellcolor[HTML]{E0F7E0}\textbf{{Configuration Issue}}} 
                                          & \cellcolor[HTML]{D9E2F3}7       & \cellcolor[HTML]{D9E2F3}6         & \cellcolor[HTML]{D9E2F3}10 
                                          & \cellcolor[HTML]{D9E2F3}6       & \cellcolor[HTML]{D9E2F3}1         & \cellcolor[HTML]{D9E2F3}3 
                                          & \cellcolor[HTML]{D9E2F3}1        \\
\multirow{1}{*}{\cellcolor[HTML]{FFFFE0}\textbf{{Network Issue}}} 
                                          & \cellcolor[HTML]{D9E2F3}4       & \cellcolor[HTML]{8EAADB}13        & \cellcolor[HTML]{D9E2F3}5 
                                          & \cellcolor[HTML]{D9E2F3}5       & \cellcolor[HTML]{D9E2F3}6         & \cellcolor[HTML]{F2F2F2}0 
                                          & \cellcolor[HTML]{F2F2F2}0       \\
\multirow{1}{*}{\cellcolor[HTML]{E0F7E0}\textbf{{Memory Issue}}} 
                                          & \cellcolor[HTML]{8EAADB}11      & \cellcolor[HTML]{D9E2F3}8         & \cellcolor[HTML]{D9E2F3}2 
                                          & \cellcolor[HTML]{D9E2F3}3       & \cellcolor[HTML]{D9E2F3}8         & \cellcolor[HTML]{D9E2F3}1 
                                          & \cellcolor[HTML]{D9E2F3}2       \\
\multirow{1}{*}{\cellcolor[HTML]{FFFFE0}\textbf{{Prompt Issue}}} 
                                          & \cellcolor[HTML]{8EAADB}16      & \cellcolor[HTML]{D9E2F3}3         & \cellcolor[HTML]{F2F2F2}0 
                                          & \cellcolor[HTML]{D9E2F3}5       & \cellcolor[HTML]{D9E2F3}3         & \cellcolor[HTML]{D9E2F3}1 
                                          & \cellcolor[HTML]{F2F2F2}0         \\
\multirow{1}{*}{\cellcolor[HTML]{E0F7E0}\textbf{{Security Issue}}} 
                                          & \cellcolor[HTML]{D9E2F3}6       & \cellcolor[HTML]{D9E2F3}2         & \cellcolor[HTML]{D9E2F3}9 
                                          & \cellcolor[HTML]{D9E2F3}2       & \cellcolor[HTML]{D9E2F3}3         & \cellcolor[HTML]{D9E2F3}4          
                                          & \cellcolor[HTML]{D9E2F3}1           \\
\multirow{1}{*}{\cellcolor[HTML]{FFFFE0}\textbf{{GUI Issue}}} 
                                          & \cellcolor[HTML]{D9E2F3}1       & \cellcolor[HTML]{D9E2F3}1         & \cellcolor[HTML]{D9E2F3}3 
                                          & \cellcolor[HTML]{8EAADB}14      & \cellcolor[HTML]{D9E2F3}1         & \cellcolor[HTML]{D9E2F3}1 
                                          & \cellcolor[HTML]{F2F2F2}0           \\
\multirow{1}{*}{\cellcolor[HTML]{E0F7E0}\textbf{{Database Issue}}} 
                                          & \cellcolor[HTML]{D9E2F3}2       & \cellcolor[HTML]{D9E2F3}4         & \cellcolor[HTML]{D9E2F3}4 
                                          & \cellcolor[HTML]{D9E2F3}10      & \cellcolor[HTML]{D9E2F3}2         & \cellcolor[HTML]{F2F2F2}0 
                                          & \cellcolor[HTML]{F2F2F2}0         \\
\end{tabular}
\end{adjustbox}
\begin{minipage}{18cm} 
\vspace{0.1cm}
\vspace{0.1cm}
  \textbf{Full names of every solution category abbreviation}: OM: \textit{Optimize Model}; AP: \textit{Adjust Parameter}; ACO: \textit{Adjust Configuration and Operation}; OC: \textit{Optimize Component}; OSRM: \textit{Optimize System Resource Management}; MD: \textit{Modify Documentation}; CHSE: \textit{Change Hardware and Software Environment}.
\end{minipage}
\end{table*}

\subsubsection{\textbf{Interpretation}}
\textbf{Frequency of the \textit{solutions}}: OM is the most frequently employed solution for addressing issues. The models are the core functional components of LLM open-source software. Therefore, OM can directly enhance the performance of models, including accuracy, response speed, and the capability to handle complex tasks. Furthermore, fine-tuning the models can ensure more stable performance across various hardware and software environments, thereby guaranteeing its effective operation under diverse conditions.
In addition, AP, ACO, and OC are also widely accepted by practitioners as solutions for problem-solving. LLM open-source software contains numerous parameters, especially in the models. Therefore, adjusting parameter settings and corresponding values, and implementing appropriate parameter passing methods, are feasible and effective approaches. 
Proper configuration and operation are essential for any software to function correctly, and LLM open-source software is no exception. Consequently, ACO is an effective approach to problem-solving. Furthermore, LLM open-source software is composed of various components, so as a general solution, OC is also widely accepted by practitioners as a method for addressing the issues. 
Although each of the remaining three categories of solutions for addressing the issues accounts for a relatively minor proportion, they still cannot be ignored. For example, OSRM reflects that practitioners can address issues within LLM open-source software by focusing on system resources.

\textbf{Mapping of \textit{solutions} to \textit{issues}}: For \textit{Code Issue}, effective solutions have already been implemented. Since issues in programming directly affect the proper functioning of LLM open-source software, especially issues related to the logic of functions. Therefore, \textit{Code Issue} tends to receive significant attention and is usually addressed promptly. 
OM constitutes a significant portion of the solution categories in every \textit{Issue} category and is the most prevalent solution category overall. It is due to the predominance of \textit{Model Issue} and \textit{Model Problem}, and LLMs are core function component of LLM open-source software. Regardless of the category of the issue, the performance of the model directly impacts the overall performance of LLM open-source software. Therefore, OM not only enhances the capability of models to address specific issues, but also improves the stability and efficiency of the entire software. 
Furthermore, \textit{Model Issue} can be addressed through all categories of solutions, which indicates that there are multiple approaches to resolving issues concerning LLMs. The resolution rate for all categories of issues has exceeded 70\%, which indicates that issues within LLM open-source software have received widespread attention from practitioners and have been effectively addressed.
%The most frequently employed solution to \textit{Model Issue} is OM, to \textit{Component Issue} is OC, to \textit{Parameter Issue} is AP, and to \textit{Documentation Issue} is MD. Generally, practitioners tend to address issues directly from their manifestations. OM constitutes a significant portion of the solution categories in every \textit{Issue} category and is the most prevalent solution category overall. It is due to the predominance of \textit{Model Issue} and \textit{Model Problem}. Furthermore, \textit{Model Issue} can be addressed through various categories of solutions, which indicates that there are multiple approaches to resolving issues related to LLMs. In addition, OM is also the primary solution for \textit{Performance Issue} and \textit{Code Issue}.

\section{Implications}
\label{sec:implication}
In this section, we discuss the implications for practitioners and researchers of LLM open-source projects based on the study results presented in Section \ref{sec:results}.

\subsection{\textcolor{black}{Implications for Users of LLM Open-Source Software}}
\begin{tcolorbox}[arc=0mm,width=\columnwidth,
                  top=1mm,left=1mm,  right=1mm, bottom=1mm,
                  boxrule=.2pt]
\textbf{Implication 1}. Users of LLM open-source software can optimize input prompts to obtain more satisfactory answers from LLM open-source software.
\end{tcolorbox}
\textit{Answer Issue} ranks the fourth among all 15 categories of issues, and within \textit{Answer Issue}, the dominant type is \textit{Poor-quality Answer} (42.7\%), which indicates a level of dissatisfaction among users with the responses provided by LLM open-source software. However, some techniques can be employed to improve the input prompts in order to obtain better responses. For example, users can utilize predefined templates to input prompts for obtaining more refined responses. A user suggested that ``\textit{you can provide context like below as in the Alpaca data}'', which indicates that users can mimic the format of training data by segmenting input prompts into multiple parts such as instructions, input, and response, and write the context according to the syntax and standards of the training data to achieve better responses (dolly \#55). Furthermore, users can also divide input prompts that exceed the length limit into multiple segments and enter these segments separately, filter out sensitive vocabulary, or provide more specific scenarios to obtain better responses. \textcolor{black}{For example, the langchain project provides several built-in text splitting tools, such as \textit{CharacterTextSplitter}, which can divide a document into smaller chunks for input into LLMs, and allows maintaining continuity between chunks by setting ``\textit{chunk\_overlap}''.}

\begin{tcolorbox}[arc=0mm,width=\columnwidth,
                  top=1mm,left=1mm,  right=1mm, bottom=1mm,
                  boxrule=.2pt]
\textcolor{black}{\textbf{Implication 2}. Users can consider using LLM Development Frameworks as assistants for their software development, while remaining mindful of the potential problems and risks associated with LLMs.}
\end{tcolorbox}
\textcolor{black}{Over half of the LLM open-source projects investigated in this study are frameworks for developing software by leveraging LLMs (see Table~\ref{Domains of Selected Projects}), which could lower the barrier for users to develop software systems with the assistance of LLMs. For example, the langchain project provides an open-source software framework for developing applications powered by LLMs, and it supports both open-source and commercial LLMs (such as GPT from OpenAI, Llama from Meta), allowing users of the langchain project to select the LLMs that fit their needs in software development. 
However, according to the results of our study, there are plenty of issues associated with generating code using LLMs. 46.3\% of Model Issues, 71.7\% of Component Issues, 65.2\% of Parameter Issues, 53.7\% of Answer Issues, and 63.4\% of Performance Issues originate from LLM Development Frameworks, which indicates that using LLM open-source software for development might lead to many potential problems and risks. For example, \cite{fu2025security} examined the security risks of Copilot-generated code, and found that that 29.5\% of Python and 24.2\% of JavaScript code snippets contain security weaknesses. Users of LLM Development Frameworks need to be aware of such issues associated with LLMs and take appropriate measures to prevent insecure code merging into the code base.}
% \textcolor{black}{Fig.~\ref{fig:The distribution of the duration of closed issues in the final sample} illustrates the distribution of durations from opening to closing for closed issues in the final sample. It is obvious that 40.34\% of the closed issues are resolved within seven days, which indicates that most issues can be resolved within a short period. However, 44.97\% of the closed issues are resolved after more than 30 days or are automatically closed. For example, in the project ``text\_generation\_webui'', closed issues that remain inactive for more than 30 days will be automatically closed. Therefore, users should promptly focus on their issues to let their issues be active. Besides, users can also add some labels (e.g., bug report) to their issues, and use a template to provide a detailed description of the issue, which can make it convenient for other developers and users to address the issue. For example, the user who raised semantic-kernel \#3797 added the ``\textit{v1 bugbash}'' label to the issue and used a template to provide a detailed description of the reproduction steps of the bug, expected behavior, operating environment, and other relevant information, along with a screenshot. The issue was ultimately resolved within four days.}

\subsection{\textcolor{black}{Implications for Developers of LLM Open-Source Software}}
\begin{tcolorbox}[arc=0mm,width=\columnwidth,
                  top=1mm,left=1mm,  right=1mm, bottom=1mm,
                  boxrule=.2pt]
\textbf{Implication 3}. Developers of LLM open-source software should possess a basic set of knowledge of LLMs.
\end{tcolorbox}
 According to our study results, the following knowledge is required to develop and use LLM open-source software: 
(1) Developers need knowledge of configuring the development and runtime environments for LLM open-source software. According to our study results, \textit{Configuration and Connection Problem} ranks the second among the causes of the issues, which indicates that practitioners need solid knowledge about environment configuration of LLM open-source software. For example, a contributor received a message that ``\textit{Package `tokenizers' not found in index and Package `taggers' not found in index}'', which indicates key packages are missing from the dependency environment when configuring the development environment for LLM open-source software (langchain \#8419). 
(2) Developers need to understand basic text processing techniques to handle input and output texts. LLMs are a kind of PLMs, and their input and output are text. Therefore, when developing LLM open-source software, it is necessary to have a basic understanding of text processing techniques like tokenization, stemming, and lemmatization, stop word removal. For example, a contributor reported that ``\textit{--no-stream is very slow, because it ignores the stop words filled into stopping\_criteria}'', which indicates that operational efficiency of LLM open-source software can be enhanced after the removal of stop words (text-generation-webui \#805). 
(3) Developers need to have a basic understanding of deep learning frameworks that LLMs (e.g., GPT4) are based on. Acquiring knowledge of deep learning frameworks can facilitate the fine-tuning and inference functionalities of LLMs. For example, a contributor received a message that ``\textit{RuntimeError: PyTorch is not linked with support for mps devices1}'', which indicates that practitioners need a basic understanding of PyTorch to deploy the software (FastChat \#854).

\begin{tcolorbox}[arc=0mm,width=\columnwidth,
                  top=1mm,left=1mm,  right=1mm, bottom=1mm,
                  boxrule=.2pt]
\textbf{Implication 4}. Developers should use more efficient parameter fine-tuning methods and design more robust parameter validation mechanisms.
\end{tcolorbox}
According to Table \ref{tab:issues-causes}, it can be observed that 22 cases of \textit{Model Issue} are caused by \textit{Parameter Problem}, 24 cases of \textit{Parameter Issue} are induced by \textit{Model Problem}, which reflects that \textit{Model Issues} and \textit{Parameter Issues} in LLM open-source software are interrelated, highlighting the critical importance of parameter settings, fine-tuning, and validation for the stability and performance of models. 
Therefore, developers need to introduce more effective parameter tuning methods for LLMs to avoid potential negative impacts on the performance of models. \cite{zhou2024empirical} suggested employing efficient parameter-efficient fine-tuning (PEFT) methods such as LoRA, IA3, Adapter, and Prefix-Tuning for fine-tuning multimodal models, which may even outperform full fine-tuning (FFT). Additionally, developers (e.g., developers of LLMs, OpenAI) should design more robust parameter validation mechanisms to minimize \textit{Model Issues} arising from incorrect parameter values and types. \textcolor{black}{For example, \cite{frantar2023sparsegpt} proposed a method called SparseGPT, which significantly compresses LLMs by pruning a vast number of parameters. The method uses Hessian information to determine which weights should be pruned and which should be retained. During the pruning process, the pruning mask and weight updates are dynamically adjusted based on the error at each layer and Hessian information to validate whether the pruned parameters negatively affect the accuracy of the LLMs.}

\begin{tcolorbox}[arc=0mm,width=\columnwidth,
                  top=1mm,left=1mm,  right=1mm, bottom=1mm,
                  boxrule=.2pt]
\textbf{Implication 5}. Developers need to be cautious when selecting models and components when developing LLM open-source software.
\end{tcolorbox}
As mentioned in Section \ref{subsec:issuesresults}, LLM open-source software contains various components, and some LLM open-source software integrates multiple models, which are developed by different teams and communities. Combining these models and components into one software system may lead to incompatibilities, such as \textsc{Model Incompatibilities} and \textsc{Component Incompatibilities}. Additionally, updates of these models and components can also lead to incompatibilities. Developers should carefully read the documentation of models and components to understand their supported features and provided interfaces. When introducing new versions of models and components in LLM open-source software, developers should carefully consider and explore the potential impact of these models and components on the compatibility within the existing system.

% \begin{tcolorbox}[arc=0mm,width=\columnwidth,
%                   top=1mm,left=1mm,  right=1mm, bottom=1mm,
%                   boxrule=.2pt]
% \textbf{Implication 6}. \textcolor{black}{Developers can consider integrating LLMs with other application domains to enhance the competitiveness of software products.
% }
% \end{tcolorbox}
% \textcolor{black}{Table~\ref{Domains of Selected Projects} indicates that most (over half) of LLM open-source projects focus on how to use LLMs to develop the software. However, LLMs can also be integrated with other application domains, and developers can identify new opportunities from cross-domain needs, driving the diversification of LLM applications. For example, ChatGPT, developed by OpenAI, has gained widespread attention worldwide and can be applied extensively in areas such as conversation, question answering, and content creation. Why not develop an open-source version of ChatGPT? Furthermore, in the 15 representative projects we selected, pandas-ai is an LLM tool used for data analysis. This indicates that developers can integrate LLMs with industries such as data analysis, healthcare, and finance to create more competitive products.}

\subsection{\textcolor{black}{Implications for Researchers of LLM Open-Source Software}}
\begin{tcolorbox}[arc=0mm,width=\columnwidth,
                  top=1mm,left=1mm,  right=1mm, bottom=1mm,
                  boxrule=.2pt]
\textbf{Implication 6}. Researchers may consider to provide automated parameter adjustment tools and methods.
\end{tcolorbox}
According to Table \ref{tab:issues-causes}, it is obvious that \textit{Model Issues} and \textit{Parameter Issues} in LLM open-source software are interrelated. Proper parameter settings significantly affect the performance of LLMs. LLM open-source software is used by plenty of end-users, many of whom may not be proficient in parameter settings. However, LLMs contain billions of parameters (e.g., learning rate, batch size, and weight decay), manual adjustments is time-wasting and inefficient. Automatic tools can systematically search for the optimal parameter settings to ensure better model performance and to make it easier for practitioners to use LLM open-source software. For example, \cite{falkner2018bohb} proposed a hyperparameter tuning method called BOHB, which combines the advantages of Bayesian optimization and bandit-based methods to achieve both stable performance and rapid convergence to optimal settings.
%Due to the randomness of LLMs, among 226 unsolved issues, 38 issues show difficulties in reproducing the problems and 30 issues present challenges in identifying the underlying causes, thereby contributing to their unresolved status. For example, after several users reported that "\textit{api\_server runs too slowly}" when chatting with the model, some users reported that ``\textit{I do not see that problem, really}'', which indicates difficulties in reproducing the problem (FastChat \#1499). Furthermore, a user reported that an issue occurred when generating answers and saving answers locally, another user encountered the same problem, but finally they reported ``\textit{not sure why}'', which indicates that sometimes the causes of the issue are difficult to locate (FastChat \#1974). 

\begin{tcolorbox}[arc=0mm,width=\columnwidth,
                  top=1mm,left=1mm,  right=1mm, bottom=1mm,
                  boxrule=.2pt]
\textbf{Implication 7}. Researchers may consider to provide methods to reproduce the problems occurred in LLM open-source software.
\end{tcolorbox}
Due to the randomness of LLMs, among 226 unsolved issues, 38 issues show difficulties in reproducing the problems. Without accurately reproducing the problem within LLM open-source software, it can be difficult for practitioners to identify root causes of the issues, thereby hindering the resolution of the issues. For example, after several users reported that ``\textit{api\_server runs too slowly}'' when chatting with the model, some other users reported that ``\textit{I do not see that problem, really}'', which reflects difficulties in reproducing the problem (FastChat \#1499), leading to the issue being unsolved.
%Agents are intelligent entities that autonomously perform specific tasks through environmental perception, strategic self-planning, and action execution \citep{he2024lma}. In LLM open-source software, the functionalities of agents involve information interaction management with models, task execution, model optimization, context maintenance, error handling, resource management and so on. In other words, the functionalities of agents almost encompass all of the software capabilities. However, according to our study results, the capabilities of agents are limited. The limitations of agent capabilities also encompass multiple aspects, such as \textit{Model Issue}, \textit{Parameter Issue}, \textit{Component Issue}, \textit{Performance Issue}. For example, a user reported that the agent ``\textit{only executes a single action on each planning step}'', which indicates that there are limitations in the agent's capability to execute tasks (langchain \#3666). Furthermore, a user reported that the agent ``\textit{could not parse LLM output}'', which indicates there are limitations in the agent's ability to interact with LLMs (langchain \#9658). In addition, users have reported functional coupling between agents, limitations in throughput of the agents and so on. Therefore, the research of approaches to expand and optimize the capabilities of agents should be a major focus of future work.

\begin{tcolorbox}[arc=0mm,width=\columnwidth,
                  top=1mm,left=1mm,  right=1mm, bottom=1mm,
                  boxrule=.2pt]
\textbf{Implication 8}. Researchers may consider establishing standardized benchmarks to evaluate the impact of different configurations (e.g., model, dataset, hardware, and software environment) on the performance of LLMs.
\end{tcolorbox}
According to Table \ref{tab:issues-causes}, \textit{Configuration and Connection Problem} is the most frequent cause to \textit{Model Issues}, and according to Table \ref{tab:issues-solutions}, \textit{Adjust Configuration and Operation} addresses 37 \textit{Model Issues} among the 193 resolved \textit{Model Issues}. Therefore, different configurations (e.g., model, dataset, hardware, and software environment) may impact the performance of models when developing and using LLM open-source software. To better understand the influences of these configurations and improve model performance, researchers can establish standardized benchmarks for evaluating the impact of various configurations on the LLM performance. These benchmarks can facilitate comparing model performance under various configurations, providing a foundation for further optimization and enhancement of the performance of LLM open-source software.

\begin{tcolorbox}[arc=0mm,width=\columnwidth,
                  top=1mm,left=1mm,  right=1mm, bottom=1mm,
                  boxrule=.2pt]
\textbf{Implication 9}. Researchers may consider exploring how to better coordinate the collaboration between models and the interoperation between components to reduce incompatibilities within the system.
\end{tcolorbox}
As mentioned in Section \ref{subsec:causesresults}, \textsc{Model Incompatibility} is the dominant type of \textit{Model Problem}, and \textsc{Component Incompatibility} is the largest type of \textit{Component Problem}. In LLM open-source software, there are various components, and some LLM open-source software integrates multiple models. These models and components are mostly developed by different teams, which may lead to incompatibilities during the integration of these models and components, the collaboration of the models to answer user questions or complete tasks, as well as the interoperation of the components to run the LLM open-source software. By coordinating collaboration between models and interoperation between components, issues induced by incompatibilities can be reduced, thereby enhancing the extensibility of LLM open-source software, allowing developers to more easily introduce new models and add new components to implement new features and extend existing functionalities.

\section{Threats to Validity}
\label{sec:threats}

%The discussion pertains to the validity threats in accordance with the directives outlined in \cite{runeson2009guidelines}. It should be noted that internal validity is not discussed, as our study did not delve into the examination of associations between variables and outcomes.

\textbf{Construct validity}: 
Due to the manual operation of data collection, labelling, sampling, and extraction in our study, there is a risk of personal bias on the data labelling and extraction results. To mitigate this risk and enhance the objectivity and reliability of our study results, we conducted pilot studies before each formal step to assess the effectiveness of methods and criteria among different researchers, which increased the construct validity of the study. Following each pilot and formal study, the first author engaged in discussions with the second and third authors to reach consensus on the data analysis results aligned with the RQs. \textcolor{black}{Since the open coding procedure was carried out by the first author, we acknowledge the potential impact of overconfidence of the first author on the validity of the data extraction results. However, the first author had five years of experiences in software development two years of experience in using LLMs, which partially mitigates the potential threat to the validity of the data extraction results. Besides, we consider adopting automated or semi-automated tools as a future measure to mitigate potential biases in data selection.}

\textbf{External validity}: 
In our study, the main concern with external validity lies in the selection of data sources. In order to maximize external validity, we chose to utilize the ``Issues'' page of GitHub projects as our data source. GitHub provides an ``Issues'' page for each open-source project, where practitioners can raise questions, report difficulties, errors, and bugs, and engage in discussions with other practitioners to address these issues. To ensure the extraction of effective information on \textit{causes} and \textit{solutions}, we opted to use closed issue information. Furthermore, GitHub, as the largest open-source project hosting platform globally, offers us sufficient research data to constitute our dataset. However, despite our considerations and the implementation of these strategies, we must admit that we may have overlooked some valuable information.

\textbf{Reliability}: 
To mitigate potential uncertainties arising from the research methodologies employed, we implemented several measures to enhance the reliability of our study. As mentioned in Section \ref{pilotdatalabelling}, we conducted pilot studies to assess consistency on data labelling results between the first and third authors before the formal studies. In the pilot data labelling phase, we computed a Cohen's Kappa coefficient value of 0.838, which indicated good consistency between authors. Throughout the process of data labelling, extraction, and analysis, thorough discussions and resolutions were conducted among the authors to address any internal inconsistencies, ensuring the consistency and accuracy of the results. Furthermore, we have provided the dataset of our study \citep{dataset} for other researchers to replicate this study and validate our findings.

\section{Related Work} \label{sec:relatedWork}

\subsection{Utilization of Pre-Trained Models in Open-Source Projects}
Several studies focused on exploring the applications of Pre-Trained Models (PTMs) in open-source projects. 
\cite{tufano2024empirical} analyzed 1,501 commits, pull requests, and issues collected from GitHub, identifying instances that suggest the use of ChatGPT in software development. They concluded that practitioners can use ChatGPT to assist in various aspects of software development tasks. 
\cite{lin2024empirical} used Latent Dirichlet Allocation (LDA) topic modeling to explore the topics of issues discussed in 200 ChatGPT-related projects, and evaluated the difficulty of resolving various categories of issues based on the average attention and closing rate of an issue category. They also analyzed the temporal trends in the evolution of these issue categories.  
\cite{pepe2024empirical} analyzed PTMs from the Hugging Face (HF) platform and identified open-source projects on GitHub utilizing these models. They concluded that it is significant to improve model transparency and documentation to foster trust in AI models. 
\cite{joel2024analyzing} collected data from the HF platform and the HFCommunity dataset, then employed neural network methods to classify submission information and conducted LDA topic modeling to analyze model card content. Their study revealed trends in popularity growth of HF and proposed a model maintenance status classification system for models based on multiple attributes to differentiate between high-maintenance and low-maintenance models. 
\cite{tan2024challenges} collected PTM-related questions from the Stack Overflow platform and analyzed the data using statistical and qualitative methods. The research revealed that PTM-related questions have become increasingly popular over time, with low response rates and longer response times. In addition, PTM-related questions were categorized during the analysis. 
\cite{taraghi2024empirical} used all public topics from the HF forum and model data from the HF Model Hub to explore the challenges users encounter when utilizing PTMs. Their study revealed new challenges in PTM reuse within the HF community and provided guidance and recommendations for improving PTM reuse for various stakeholders. 
\cite{tao2024empirical} conducted an empirical study to analyze challenges faced by LLMs in addressing GitHub issues, identifying factors influencing their performance. They subsequently introduced a novel multi-agent framework called MAGIS, which collaboratively addresses GitHub issues. MAGIS demonstrates significant performance improvements over baseline models and other popular LLMs, achieving an eight-fold improvement in resolution rate compared to the application of GPT-4.

\subsection{Exploration of Issues, Causes, and Solutions in Open Source Projects}
Several studies used automatic and manual methods to conduct empirical software engineering research on open-source software projects. 
\cite{zhou2023airepo} selected 576 repositories from the PapersWithCode platform, then collected 24,953 issues from these repositories. They manually analyzed these issues and categorized them. They then examined the resolution status of the issues and explored the use of GitHub issue management features (such as labels and assignees) and their impact on issue resolution. Ultimately, they classified the issues into 13 categories. The two most common issues are runtime errors (23.18\%) and unclear instructions (19.53\%). Among these, 67.5\% of the issues were closed, with half of them resolved within four days. The GitHub issue management features are not widely adopted in open-source AI repositories.
\cite{humbatova2019taxonomy} employed the open coding approach to manually analyze 1,059 artifacts obtained from GitHub issues, pull requests, commits, and related Stack Overflow posts. Additionally, they collected failure experiences encountered by researchers and practitioners during the use of deep learning frameworks through interviews with 20 individuals. Ultimately, they developed a classification system consisting of 5 top-level categories, covering a range of failure types from model architecture to the model training. They validated this classification system through a survey involving 21 developers, confirming that almost all failure categories (13/15) were experienced by at least 50\% of the survey participants.
\cite{martineztorres2014ossc} selected 154 papers published between 2001 and 2011 from the Web of Science database using “open source” and “virtual/online communities” as searching keywords. Subsequently, they performed cluster analysis on the keywords of these papers. Finally, they identified eight distinct research themes, and discussed the main challenges and academic achievements for each category.
\cite{zhao2023performance} collected a total of 570 performance issues from 13 open source projects from the Apache Software Foundation platform. Then, they employed the open coding approach to manually analyze these issues. Finally they summarized eight general types of performance issues with corresponding root causes and resolutions that apply for three languages (i.e., C/C++, Java, Python). They found that only 15\% of the performance issues involved revisions to test code, and design-level optimizations typically require greater investment, but do not always lead to higher performance gains.
\cite{zahedi2018empirical} collected the issues from 200
randomly sampled GitHub repositories, and used a mixed-method approach, combining LDA topic modeling and manual thematic analysis, to conduct the study. Finally, they summarized 7 high-level themes of problems that developers face in implementing security features.
\cite{waseem2021empirical} collected 1,345 issue discussions from five microservice systems hosted on GitHub and manually  analyzed these issues. Finally they developed a classification system for microservices issues, consisting of 17 categories, 46 subcategories, and 137 issue types, and identified seven categories of root causes for the issues.
\cite{waseem2023empirical} collected data from 2,641 issues from the issue tracking systems of 15 open-source microservice systems on GitHub, conducted 15 interviews, and carried out an online survey completed by 150 practitioners. They used descriptive statistics and constant comparison techniques to analyze the data. Finally, they developed a classification system for microservice issues consisting of 19 categories, 54 subcategories, and 402 types; a classification system for root causes consisting of 8 categories, 26 subcategories, and 228 types; and a classification system for solutions consisting of 8 categories, 32 subcategories, and 177 types.

\subsection{Conclusive Summary}
As LLMs are increasingly used in our daily life, many open-source software practitioners started to integrate LLMs into open-source software to implement core functionalities \citep{weber2024large}. In the field of open-source software development, many researchers have already employed automated and manual methods to conduct empirical studies on specific types of open-source projects. However, existing studies did not focus on investigating the issues faced by LLM open-source software from the perspective of practitioners (i.e., developers and users), nor did they explore the underlying causes of these issues and their solutions.
%Based on the existing research, it is evident that researchers show significant interest in LLMs. However, most studies focus either on the application and performance of LLMs in SE tasks or on the design of LLMs specifically tailored for a certain stage of software development. 
\cite{lin2024empirical} had studied the issues within ChatGPT-related projects, which is a specific and popular commercial LLM, and they utilized the LDA topic modeling, which is an automatic method, to explore the topics of the issues discussed in 200 ChatGPT-related projects, but they did not explore the causes and solutions of these issues. %Moreover, their taxonomy of issues only provides category names and the top ten keywords under each category, whereas our taxonomy offers a two-tier classification, including categories and all corresponding types under each category.
Our study, grounded in the perspective of practitioners, aims to explore the issues within LLM open-source projects, their underlying causes, and potential solutions. We employed a manual qualitative approach to extract and analyze the data collected in the study. To ensure the representativeness of our research results, we collected projects from GitHub, the largest open-source project hosting platform in the world.
Ultimately, \cite{lin2024empirical} provided a one-tier classification of issues, comprising a total of ten categories, with the top ten keywords for each category identified by the automatic method. However, our taxonomy offers a two-tier classification of issues, comprising a total of 15 categories with corresponding types under each category. We found that only the issue category \textit{Model Reply} in the issue taxonomy identified by \cite{lin2024empirical} is similar to the issue category \textit{Answer Issue} in our classification. The other categories in their issue taxonomy are quite different from our results. %Moreover, we also provided a two-tier classification for causes and solutions. 

\section{Conclusions} \label{sec:conclusions}
In this study, we focused on the issues that practitioners encounter when developing and using LLM open-source software, as well as their underlying causes and potential solutions. We collected all closed issues from LLM open-source projects on GitHub that met our criteria and labelled a total of 14,476 closed issues that satisfied our requirements. Then we randomly chosen 994 closed issues from the labelled closed issues as the sample data of our study. Finally, we extracted a total of 994 issues, 559 causes, and 798 solutions based on our data extraction criteria. The results show that \textit{Model Issue} is the most common issue faced by practitioners, \textit{Model Problem}, \textit{Configuration and Connection Problem}, and \textit{Feature and Method Problem} are identified as the most frequent causes of the issues, and \textit{Optimize Model} is the predominant solution to the issues. 

Based on the study results, we provide implications for practitioners and researchers of LLM open-source projects: 
Due to the interrelated effects of \textit{Model Issues} and \textit{Parameter Issues} within the software, practitioners should pay more attention to the proper settings of parameters and the validation mechanisms for these parameters. In addition, users can optimize prompts (e.g., constructing prompts according to training data templates) to obtain more satisfactory responses from LLM open-source software. Developers need to be cautious to select suitable models and components when developing LLM open-source software. 
Besides, researchers should provide automated parameter adjustment tools and methods to ensure better model performance in LLM open-source software. Moreover, researchers should explore potential solutions for reproducing the problems to address those issues that remain unsolved. Researchers can also consider to provide standardized benchmarks to evaluate the impact of different configurations on the performance of models.

\section*{Data availability}
We have shared the link to our dataset in the reference \citep{dataset}. 

\section*{Acknowledgments}
This work is supported by the National Natural Science Foundation of China under Grant No. 62172311 and the Major Science and Technology Project of Hubei Province under Grant No. 2024BAA008.

\printcredits

\bibliographystyle{cas-model2-names}
\bibliography{references}
\balance
\end{sloppypar}
\end{document}